\documentclass[10pt]{article}

\usepackage[margin=1in]{geometry}

\usepackage{amsmath}
\usepackage{graphicx}
\usepackage{enumerate}
\usepackage{natbib}
\usepackage{url} 
\usepackage{bbm}
\usepackage{amssymb}
\usepackage[boxed,linesnumbered]{algorithm2e}
\usepackage{amsthm}
\usepackage{enumerate}
\usepackage{subcaption}
\usepackage{appendix}
\usepackage[colorlinks,allcolors=black]{hyperref}

\newcommand{\Rho}{\mathrm{P}} 
\DeclareMathOperator{\diag}{diag}

\SetAlgorithmName{Procedure}{procedure}{List of Procedures}

\title{Prior Elicitation for Generalised Linear Models and Extensions}
\author{Geoffrey R. Hosack$^1$}
\date{%
	$^1$CSIRO--Data61\\
	Hobart, Tasmania 7001, Australia \\
	E-mail: {geoff.hosack@csiro.au}
}

\begin{document}

\maketitle

\begin{abstract}
A statistical method for the elicitation of priors in Bayesian generalised linear models (GLMs) and extensions is proposed. Probabilistic predictions are elicited from the expert to parametrise a multivariate t prior distribution for the unknown linear coefficients of the GLM and an inverse gamma prior for the dispersion parameter, if unknown. The elicited predictions condition on defined elicitation scenarios. Dependencies among scenarios are then elicited from the expert by additionally conditioning on hypothetical experiments. Elicited conditional medians efficiently parametrise a canonical vine copula model of dependence that may be truncated for efficiency. The statistical elicitation method permits prior parametrisation of GLMs with alternative choices of design matrices or observation models from the same elicitation session. Extensions of the method apply to multivariate data, data with bounded support, semi-continuous data with point mass at zero, and count data with overdispersion or zero-inflation. A case study elicits a prior for an extended GLM embedded in a statistical model of overdispersed counts described by a binomial-simplex mixture distribution. The elicited canonical vine model of dependence is found to incorporate substantial information into the prior. The procedures of the statistical elicitation method are implemented in the \textsf{R} package \textbf{eglm}.
\end{abstract}

{
	\hypersetup{linkcolor=black}
}

\section{Introduction}\label{sec:introduction}

Generalised linear models \citep[GLMs;][]{Nelder1972, McCullagh1989} provide foundational support for applied statistical inference in science, engineering and medicine. To begin the process of Bayesian inference for the unknown parameters of a GLM, a prior must first be specified. This first inferential step can entail a statistical and scientific investigation similar to an empirically focused study \citep{Garthwaite2005, Kuhnert2011}. A general and structured elicitation procedure that supports Bayesian GLMs and related models is therefore needed.  The statistical method for elicitation presented here is applicable to GLMs that allow for a range of response data with either known or unknown dispersion. Moreover, the elicitation method accommodates extensions to statistical models for multivariate, bounded support, semi-continuous, overdispersed or zero-inflated  data. 

GLMs combine random and systematic components to describe data \citep[][Section \ref{sec:glm}]{Nelder1972}. Elicitation methods for GLMs are often based on techniques originally developed for the normal linear model (Section \ref{sec:normal}). General approaches to elicit unknown dispersion parameters, however, are unavailable for non-normal GLMs (Section \ref{sec:glm_dispersion}). Comparatively more effort has been placed on elicitation of the unknown linear coefficients of GLMs. The present approach for elicitation generalises the concept of an independent conditional means prior for the linear coefficients \citep[ICMP;][]{Bedrick1996}. An ICMP uses an assumption of conditional independence given the covariates to ease the complexity of the elicitation process (Section \ref{sec:glm-icmp}). Statistical methods of elicitation for GLMs have been proposed that relax the  assumption of conditional independence among elicited responses (Section \ref{sec:dCMPs}). These elicitation methods each emphasise most or all of the elicitation principles summarised by \citet{Kadane1998} for predictive elicitation: 1) expert opinion is worthwhile to elicit; 2) only observable quantities should be assessed via an indirect rather than direct elicitation of unknown model parameters; 3) experts should be asked for quantiles or probabilities and never be asked to estimate moments greater than the mean; 4) feedback should be provided to the expert; and 5) assessments should be provided conditionally and unconditionally on hypothetical data. 

The statistical elicitation method outlined in Section \ref{sec:specs} also incorporates these same principles, and additionally the following. By design the method is \emph{justifiable}.  The expert builds a probability model that re-expresses the elicited probability assessments without loss of information. All elicited quantities are univariate because experts typically do not accurately assess joint probabilities \citep{Garthwaite2005}. Graphical and numerical feedback mechanisms \citep[e.g.,][]{OHagan2006book} are provided that support the expert's ability to learn, adjust and critique assumptions during the elicitation session.

The method is also \emph{sequential}. The elicitation first addresses the dispersion parameter that may be known or unknown (Section \ref{sec:unknown_dispersion}). Given this characterisation, marginal elicitations for the mean responses proceed analogously to an ICMP (Section \ref{sec:marginal}). Dependence in the elicited mean responses is then modelled by a canonical vine copula \citep[][see Section \ref{sec:DCMP}]{Bedford2001, Joe2015} that efficiently captures expert opinion (Section \ref{sec:estimation}). This approach avoids asking the expert to re-evaluate central credible intervals conditionally, which is a more difficult task compared to the elicitation of marginal credible intervals \citep{Al-Awadhi1998, Al-Awadhi2001, Elfadaly2020} and requires additional elicited quantiles, but rather elicits only conditional medians instead. These elicitations condition on hypothetical data that are sequentially introduced (Section \ref{sec:conditioning}). Once introduced, the hypothetical data are retained to ease the cognitive burden on the expert \citep[sensu][]{Kadane1980}. Moreover, the sequential introduction of hypothetical data supports truncated canonical vine dependence models that further lessen the elicitation burden (Section \ref{sec:truncated-cmps}). Truncated vines are helpful in a situation where adding further information into the conditioning set of hypothetical response data has no meaningful effect on the beliefs or predictions of the expert, or where time and resource constraints limit the full exploration of dependencies. 

An additional design principle is \emph{generality}. The statistical elicitation method not only applies to a broad class of models \citep[sensu][]{Kadane1998, Mikkola2021}, but also the same elicitation session can be used to parametrise alternative GLMs.  At the time of the elicitation steps described by Sections \ref{sec:unknown_dispersion} and \ref{sec:systematic}, it is sufficient to have specified only the mean domain, link function and the elicitation scenarios for consideration by the expert. Alternative model matrices or observation models may be specified after the elicitation session concludes to induce alternative prior specifications for GLMs (Section \ref{sec:induced-prior}). The generality of the statistical method for elicitation is further illustrated in Section \ref{sec:extensions} with extensions to multivariate,  semi-continuous, bounded support, zero-inflated and overdispersed data. 

An elicitation case study demonstrates an overdispersed binomial model in Section \ref{sec:study}, where the elicited quantity is modelled with an extended GLM. The conclusion summarises future directions (Section \ref{sec:discussion}), such as the development of open and free software that is an important aspect of statistical methods for elicitation \citep{Kadane1980, Garthwaite2005, Mikkola2021}. The freely available \textsf{R} package \textbf{eglm} \citep{Hosack2023} implements the procedures of the statistical elicitation method.

\section{Review of GLMs and Elicitation Methods}\label{sec:GLMlit}

\subsection{Generalised Linear Models}\label{sec:glm}

The Generalised Linear Model (GLM) extends regression analysis developed for the normal linear model to the case where observables follow an exponential dispersion model \citep{Jorgensen1997}.  An exponential dispersion model for a random variable $\psi$ has the density function
\begin{equation}
	f(\psi \mid \theta, \lambda) = a(\psi, \phi)\exp\left\{-\lambda [\psi \theta - b(\theta)] \right\}\label{eq:EDM}
\end{equation}
for functions $a(\cdot)$ and $b(\cdot)$. The mean is $\mathbb{E}\left[\psi\right]  = \mu \in \Omega \subseteq \mathbb{R}$, where $\mu = \tau(\theta) = db(\theta)/d\theta$. The variance is $\mathbb{V}\left[\psi\right]  = \phi v(\mu)$, where $\phi = 1/\lambda > 0$ is the dispersion parameter and $v(\mu) = d^2b(\tau^{-1}(\mu))/d\mu^2$ is the variance function.  An exponential dispersion model with mean $\mu$ and dispersion parameter $\phi$ is denoted by $\psi \sim ED(\mu, \phi)$.

If the dispersion parameter is known then Eq. \eqref{eq:EDM} corresponds to a natural exponential family model \citep[][Ch. 2]{Morris1982, Jorgensen1997}. If the model is closed with respect to scale transformation such that $cED(\mu, \phi) = ED(c\mu, c^{2 - p}\phi)$ for $c > 0$ given {power} parameter $p\in (-\infty, 0] \cup [1, \infty)$ with variance function $v(\mu) = \mu^p$, then Eq. \eqref{eq:EDM} corresponds to a Tweedie exponential dispersion model \citep{Jorgensen1987}. The three standard continuous observation models used in GLMs \citep[e.g.,][]{Lindsey1997} are special cases: normal  ($p = 0$),  gamma  ($p = 2$) and inverse Gaussian  ($p = 3$). The discrete Poisson distribution is also a Tweedie model ($p = 1$), as are the semi-continuous compound Poisson distributions with point mass at zero ($1 < p < 2$).  

A GLM \citep[e.g.,][]{McCullagh1989} is defined by:
\begin{enumerate}
	\item The observations $\psi_{j(i)} \sim ED(\mu_i, \phi_i)$ for $i = 1, \ldots, n$, $j(i) = 1, \ldots, J(i)$ and $\phi_i = \phi/w_i$ with known weight $w_i$.
	\item The linear predictor $\eta_i = x_i^\top\beta + o_i$, $i = 1, \ldots, n \geq p$ with $p$-dimensional vector $\beta$ of unknown parameters. The $p$-dimensional vector of covariates $x_i^\top$ forms the $i$\textsuperscript{th} design point, which is the $i$\textsuperscript{th} row of the $n\times p$ model matrix $X$ of full column rank. The offset $o_i$ for the linear predictor at the $i$\textsuperscript{th} design point is assumed known.
	\item A continuous link function $g$ invertible over the mean domain $\Omega$ such that $g(\mu) = \eta$. 
\end{enumerate}
The GLM consists of two basic components \citep{Nelder1972}: The ``systematic component''  defined by $\mu = \mathbb{E}\left[\psi \mid X, \beta, \phi\right] = g^{-1 }(X\beta)$ with $\beta$ unknown, and the ``random component'' that describes the uncertainty in $\psi$ conditional on $\mu$ and the dispersion parameter $\phi$.

\subsection{Elicitation for the Normal Linear Model}\label{sec:normal}

Elicitation procedures developed for the normal linear model have had a major impact on the development of probabilistic elicitation methods for GLMs \citep[][]{Garthwaite2005}. An elicitation approach for the special case of a normal linear regression model with unknown variance was initially proposed by \citet{Kadane1980}.  The procedure occurs in two stages. In the first stage, marginal quantiles are elicited at each design point defined by the covariates that correspond to a hypothetical observation. The mean is estimated by ordinary least squares and the degrees of freedom parameter then estimated by averaging over a function of the elicited quantiles evaluated at each design point. In the second stage,  conditional medians and quartiles are elicited at each of the design points given hypothetical observations. These results are then used to obtain an averaged estimate of the scale matrix for $\beta$ \citep[see][for details]{Kadane1980, Kadane1998}.  

Various alternatives to the method of \citet{Kadane1980} have been suggested. \citet{Al-Awadhi1998} note that unconditional estimates of central credible intervals (spreads) should be more accurate than conditional estimates, and propose using the method of \citet{Kadane1980} to generate preliminary  estimates of correlations that are then rescaled by marginally estimated spreads.  \citet{Al-Awadhi1998} also investigate an alternative method that avoids elicitation of conditional spreads altogether, but which does not ensure that the elicited covariance matrix is positive definite. \citet{Cooney2023} note that the method of \citet{Kadane1980} requires a large number of elicited quantiles to parametrise a multivariate normal. Their suggested approach asks fewer questions of an expert by eliciting concordance probabilities for pairwise jointly distributed unknowns to parametrise partial correlations. The elicited concordance probabilities are not guaranteed to conform with a valid correlation matrix in high dimensions. \citet{Garthwaite1988} describe an alternative method that depends on the choice of design points made by the expert. This approach limits application to regression problems with independent design points, and so prohibits for example polynomial regression or models that require factor coding.

\subsection{Unknown Dispersion Parameters in GLMs}\label{sec:glm_dispersion}

Probabilistic elicitation of unknown dispersion parameters within GLMs are uncommon outside of the normal linear model. For a GLM with a gamma observation model, \citet{Elfadaly2015} elicit an unknown dispersion parameter conditional on an assumed mean value   for $\mu$. This elicitation approach relies on a scaling property of the gamma distribution with respect to its mean, which arises as a special case of the Tweedie class (Section \ref{sec:glm}), such that $cED(\mu, \phi) = ED(c\mu, c^{2 - p}\phi) = ED(c\mu, \phi)$ for $p = 2$ and $c > 0$. A prior for the ratio of a hypothetical observation $y$ relative to the mean $\mu$ is then parametrised conditional on a given value for $\mu$. However, this scaling property available for the gamma distribution does not generalise to other exponential dispersion models commonly used in GLMs. 

\subsection{Independent Conditional Means Priors for GLMs}\label{sec:glm-icmp}

\citet{Bedrick1996} describe an approach to elicit the coefficients of the linear predictor  for GLMs within the natural exponential family, or equivalently, where the observation distribution is described by an exponential dispersion model with known dispersion parameter $\phi$. Subjective probability distributions are elicited for the mean of the response $\mu_i$ conditional on a set of covariates $x_i^\top$ for $i = 1, \ldots, n$. In its most basic formulation, the $n\times p$ model matrix $X$ for the elicitation session is full rank with $n=p$, and chosen so that the elicitations for $\mu$ may be thought of as conditionally independent given $X$. \citet{Bedrick1996} demonstrate how a conditional independence assumption for the elicited distribution of the unknown means can be reasonable, and also suggest guidelines for the choice of $X$ such that the independence assumption is credible.  Given a link function $g$ with linear model $g(\mu) = X\beta$, the induced prior for $\beta$  is known as an independent conditional means prior (ICMP). A normal approximation of the prior for $\beta$ can be a plausible assumption for ICMPs \citep{Bedrick1996}. \citet{Hosack2021} apply an ICMP approach to elicit multivariate normal priors for the unknown coefficients $\beta$ in GLMs.  \citet{Denham2007} describe an approach for logistic regression that is related to ICMPs, where Monte Carlo approximation is used to derive a prior based on samples drawn from elicited beta distribution that are assumed independent conditional on the model matrix $X$. 

\subsection{GLMs with Dependent Conditional Means Priors}\label{sec:dCMPs}

The assumption of independence among elicited responses within an ICMP approach may be overly restrictive for GLM priors. Conditional on $\phi$ and $X$, \citet{Chen2003} describe an elicitation approach that elicits a mean response and an associated degree of belief to parametrise a prior for $\beta$. Based on asymptotic results for this prior conditional on an imaginary sample, \citet{Bove2011} derive a class of multivariate normal $g$-priors and also review other related normal $g$-priors proposed for GLMs. For both the prior of \citet{Chen2003} and a normal $g$-prior, the modelled a priori dependence among design points (conditional means) is entirely determined by the choice of model matrix. \citet[][Ch. 4]{Robert2007} notes that for this reason $g$-priors may usefully substitute for prior information where such information is sparse. However, the above priors do not elicit dependent conditional mean priors in GLMs and are not further evaluated here. 

Recent approaches have elicited dependent conditional mean responses in GLMs. An elicitation method for piecewise-linear and non-normal GLMs developed by \citet{Garthwaite2013} allows for limited dependence among conditional means given a normal prior specification for $\beta$. The model structure is restricted to assume no interactions among covariates, and the elicited covariance matrix of the multivariate normal prior for $\beta$ is assumed block-diagonal. For multinomial GLMs, \citet{Elfadaly2020} adapt the elicitation algorithm of \citet{Kadane1980} in Section \ref{sec:normal} to parametrise a normal prior for $\beta$. First marginal medians are elicited for category odds, then conditional central credible intervals are elicited in a second stage. Although not working with covariates, \citet{Wilson2018} propose the use of $D$-vine copulas \citep[][Ch. 3]{Joe2015} to elicit priors for the multinomial distribution. The $D$-vine copula is related to the canonical vine representation of dependency that will be introduced in Section \ref{sec:systematic}. Unlike the $D$-vine copula, however, it will be seen that the canonical vine copula sequentially adds new conditioning information into an elicitation session in such a way so that the expert is never asked to forget information once introduced into the elicitation session.

\section{Statistical Elicitation Method for GLMs}\label{sec:specs}

\subsection{Definitions for Elicitation}\label{sec:components}

The \textit{expert} has the domain knowledge elicited by the session \textit{facilitator}  \citep{Garthwaite2005} according to a model specified by the \textit{statistician}. Each of these roles may be fulfilled by the same individual, or by different individuals or by some combination thereof. More than one individual may contribute to each role. \citet{Garthwaite2005} and \citet{OHagan2006book} provide guidance on preparation for an elicitation session. In particular, the facilitator provides the expert with probabilistic training before the elicitation. This training incorporates well-established statistical concepts that may not be fully appreciated by all domain experts. The definition of quantiles (or percentiles or fractiles) and their use in probability assessments is particularly important. Often medians are used to assess location while inter-quartile or inter-tertile distances assess spread. Another important aspect of the statistical training is education on the sensitivity of statistical estimators to sample size. Without proper training, experts can either  overemphasise \citep{Tversky1974} or underemphasise \citep{Garthwaite2005} the information contained within a sample for a summary statistic such as the sample mean. Education for an expert unfamiliar with the Central Limit Theorem (CLT), for example, may include simulated examples \citep[e.g.,][]{Dinov2008, Zhang2022} as part of the expert training.

In the structured elicitation session, the expert is asked by the facilitator to envision or predict hypothetical realisations of the \textit{elicitation target} $\psi$ that arises from a GLM (Section \ref{sec:glm}). The facilitator ensures that the elicitation target is relevant and interpretable to the expert. The elicitation session is supported by \textit{elicitation scenarios} that are conceptually defined by a $n\times p'$  matrix $U$ for the expert. The $i$\textsuperscript{th} row of $U$ defines the $i$\textsuperscript{th} scenario, $u_i^\top$, that is a $p'\times 1$ vector of measurable covariates. Elicitation scenarios  may be described numerically or by visualisations that might, for example, relate to geographical covariates \citep[e.g.,][]{Denham2007, Hosack2017}.  The facilitator ensures that the scenarios are relevant and also numerically or graphically interpretable to the expert. 

The statistician describes the elicitation target  by a set of exponential dispersion models $\mathcal{F}$ with shared mean domain $\Omega$ that is consistent with the choice of link function mapping $g: \Omega \rightarrow \mathbb{R}$. The set of exponential dispersion models $\mathcal{F}$ allows for distributions with different variance functions at the time of elicitation, if desired. The systematic component of the GLM depends on the model matrix $X$ that is linked to the elicitation scenarios of $U$. The design point $x_i^\top$, which is the $i$\textsuperscript{th} row of $X$, is a function of the covariates that define the $i$\textsuperscript{th} elicitation scenario $u_i^\top$ for $i = 1:n$. The expert is thus shielded from choices of factor coding, covariate scaling or other modification to $X$ made by the statistician that should not affect the elicitation.  Many alternative choices of $X$ with full column rank equal to $p \leq n$ may be consistent with $U$. The chosen set of these alternative model matrices is denoted by $\mathcal{X}$.

There is \textit{process uncertainty} about how the systematic component $\mu$ relates to the  scenarios of $U$ through the covariates $X$ as determined by the unknown $\beta$. The random component  accounts for \textit{observation uncertainty}, induced by sampling variability or measurement error or experimental error, that is unexplained by the covariates. The observation uncertainty is described by an exponential dispersion model (Section \ref{sec:glm}), where the dispersion parameter $\phi$ may be known or unknown. The observation uncertainty of a sample mean $\bar{\psi} = (1/w)\sum_{j = 1}^w \psi_j$, given $\mu_i$ and $\phi$ at scenario  $u_i^\top$, is reduced with an increasing number of samples $w$ (see Section \ref{sec:dispersion_EDM} for use of this property). \citet{Garthwaite1988} describe how sequential elicitation of observation and process uncertainty in the normal linear model eases the assessment tasks for experts, where the systematic component is described as the ``long run mean''. For GLMs, an exponential dispersion model has expectation $\mu_i$ at scenario $u_i^\top$, and so the sample mean $\bar{\psi}$ at scenario $u_i^\top$ converges by the law of large numbers to the systematic component $\mu_i$ as $w \rightarrow \infty$. Observation uncertainty introduced by the random component is therefore absent from the systematic component, and experts need not correct for observation uncertainty in their assessments of the systematic component.

\subsection{Model Structure for Elicitation}\label{sec:model_structure}

Let $\psi$ denote a quantity of interest that arises from a GLM (Section \ref{sec:glm}), and  $k:l = \{k, k + 1, \ldots, l\}$ denote a sequence of integers. The elicitation is structured by the prior for GLMs proposed by \citet{West1985},
\begin{subequations}\label{eq:model}
	\begin{gather}
		\psi_{j(i)} \mid \mu_i, \phi \sim f(\mu_i, \phi )\,,\quad  i = 1:n\,, \quad j(i) = 1:J(i)\,,\label{eq:model_obs}\\
		g(\mu) = \eta = X\beta + o\,,\label{eq:model_link}\\
		\beta \mid \lambda \sim N(\delta, \Sigma / \lambda)\,,\label{eq:model_beta}\\
		1/\phi = \lambda \sim G(s/2, r/2)\,,\label{eq:model_lambda}
	\end{gather}
\end{subequations}
where $N(m, S)$ is a normal distribution with mean $m$ and positive definite covariance matrix $S$,  and $G(x \mid a, b) = b^{a}x^{a - 1}\exp[- bx]/\Gamma(a)$ is a gamma distribution. 

Eq. \eqref{eq:model_obs} describes the observation uncertainty of the GLM, where $f$ is an exponential dispersion model, and Eq. \eqref{eq:model_link} defines the linear model with $n \times p$ covariate matrix $X$ of rank $p \leq n$ (Section \ref{sec:glm}). Eq. \eqref{eq:model_beta} describes the process uncertainty by a conditional normal prior for $\beta$. GLM prior specifications often use a multivariate normal for $\beta$ given $\phi$ \citep{OHagan2004}. The normal  prior requires specification of only the first two moments, and it is computationally convenient given a normal approximation of the likelihood \citep{Gelman2014}. It is sometimes also supported by asymptotic arguments, as in some $g$-prior settings (see Section \ref{sec:dCMPs}). Eq. \eqref{eq:model_lambda} is the prior for the index parameter $\lambda = 1/\phi$ that is, in general, gamma distributed. The case of known dispersion is also accommodated, as described below. 

First, if the dispersion parameter is unknown then $s, r \in (0, \infty)$ in Eq. \eqref{eq:model_lambda} and $\lambda$ is gamma distributed. The marginal prior distribution for $\beta$ is then generalised multivariate t \citep[][Ch. 5]{Kotz2004} with location  $\delta$ and scale  $\Sigma$,
\begin{equation}
	St_n(\beta \mid \delta, \Sigma, r, s) =  \frac{\Gamma((s + n)/2)}{\Gamma(s/2)(\pi r)^{n/2}}\left|\Sigma\right|^{-\frac{1}{2}}\left[1 + \frac{1}{r}(\beta - \delta)^\top \Sigma^{-1}\left(\beta - \delta\right)\right]^{-\frac{s + n}{2}}.\label{eq:beta_marginal}
\end{equation}
Although a  normal prior for $\beta$ often adequately captures prior information, \citet{West1985} and \citet{OHagan2004} also note that a multivariate t prior may be preferred because it allows for extreme values of $\beta$. Rescaling $\Sigma$ by $\mathbb{E}[\lambda] = s/r$ in Eq. \eqref{eq:beta_marginal} obtains the equivalent  multivariate t distribution denoted by $\beta \sim t_n(\delta, (r/s)\Sigma, s) = St_n(\delta, (r/s)\Sigma, s, s)$, see Section \ref{sec:t} for details. 

Second, if the dispersion parameter is known then let $s, r \rightarrow \infty$ subject to $s/r = 1/\phi$.   
In this limit, the induced prior for $\beta$ of Eq. \eqref{eq:beta_marginal} corresponds to Eq. \eqref{eq:model_beta} with $\lambda = s/r$.

\subsection{Outline of Statistical Elicitation Method}\label{sec:elicitation-protocol} 

The structured elicitation method is summarised in Procedure \ref{protocol}. Given the choices of $g$, $U$, and $\mathcal{F}$, the elicitation procedure begins by selecting, without loss of generality, an exponential dispersion model $f \in \mathcal{F}$ with variance function $v(\mu)$. The elicitation addresses the observation uncertainty of the random component in Section \ref{sec:unknown_dispersion}. The process uncertainty of the systematic component is addressed in Section \ref{sec:systematic}. During the elicitation session, the facilitator conditions the expert assessments for the systematic component on $U$ rather than $X$. After the elicitation session, the statistician can then explore alternative choices of $X \in \mathcal{X}$ consistent with $U$. To ensure this flexibility, during the elicitation session the model matrix $X$ is specified as a square full rank matrix. This non-singular transformation, given the invertible link function $g$, ensures no loss of information between the elicited systematic component $\mu \mid U$ and the induced prior on $\beta$ (Section \ref{sec:induced-prior}). 

Without loss of generality, the elicitation session in Section \ref{sec:systematic} further assumes the saturated model $X = I$ in Eq. \eqref{eq:model_link}.  In a slight abuse of notation, a known offset, if present, is absorbed into the linear predictor $\eta$ such that  $\eta = X\beta$ and $o = 0$ in Eq. \eqref{eq:model_link}. The resulting parametrisation for the linear predictor induced by Eqs. \eqref{eq:model_link} and \eqref{eq:model_beta} is 
\begin{equation}
	\eta \mid X = I, \lambda \sim N(m, V/\lambda)\,,\quad V = \diag[V]^{1/2}R\diag[V]^{1/2}\,,\label{eq:eta}
\end{equation}
where $\diag[V]$ is a diagonal matrix formed from the main diagonal entries of positive definite matrix $V$ that has positive definite correlation matrix $R$. The marginal elicitation of the systematic component described by $m$ and $\diag[V]$ proceeds in Section \ref{sec:marginal}. Elicitation of the correlation matrix $R$ progresses by means of a canonical vine copula (Section \ref{sec:DCMP}) as described in Sections \ref{sec:estimation}, \ref{sec:conditioning} and \ref{sec:truncated-cmps}. 

\begin{algorithm}[h]
	\SetAlgoLined
	\KwData{$\mathcal{F}$, $g$, $U$, $\mathcal{X}$}
	\KwResult{Estimates $\hat{\delta}$ and $\hat{\Sigma}$, Eq. \eqref{eq:model_beta}; estimates $\hat{s}$ and $\hat{r}$, Eq. \eqref{eq:model_lambda}}
	\For{Random component given  $f \in \mathcal{F}$}{
		Elicit $s$ and $r$ in Eq. \eqref{eq:model_lambda} (Section \ref{sec:unknown_dispersion} and Procedure \ref{alg:dispersion})\;
	}
	\For{Systematic component given  $g$ and $U$}{
		Elicit $m$ and $\diag[V]$ in Eq. \eqref{eq:eta} (Section \ref{sec:marginal} and Procedure \ref{alg:CI})\;
		Elicit $R$ in Eq. \eqref{eq:eta} (Sections \ref{sec:DCMP}--\ref{sec:truncated-cmps} and Procedure \ref{alg:cond})\;
	}
	Parametrise $\hat{s}$, $\hat{r}$, $\hat{\delta}$ and $\hat{\Sigma}$ given  $f'\in\mathcal{F}$ and $X \in \mathcal{X}$ (Eq. \eqref{eq:parms} in Section \ref{sec:induced-prior})\;
	Return $\hat{s}$, $\hat{r}$, $\hat{\delta}$, $\hat{\Sigma}$
	\caption{Overview of elicitation and parametrisation procedures.}\label{protocol}
\end{algorithm}

After the elicitation session concludes, the priors of Eqs. \eqref{eq:model_beta} and \eqref{eq:model_lambda} are parametrised (Section  \ref{sec:induced-prior}). First, the choice of an exponential dispersion model $f' \in \mathcal{F}$ parametrises the prior for $\lambda$ in Eq. \eqref{eq:model_lambda}. Second, the choice of model matrix $X \in \mathcal{X}$  parametrises the prior for $\beta \mid \lambda$ in Eq. \eqref{eq:model_beta}. Choices of $X \in \mathcal{X}$ with either $p = n$ or $p < n$ are supported, although the latter choice necessarily induces information loss in the parametrised prior. Extensions of the statistical elicitation method are presented in Section \ref{sec:extensions}. The method is illustrated by application that uses an extended GLM in Section \ref{sec:study}.

\section{Random Component}\label{sec:unknown_dispersion}

The random component of a GLM captures sources of observation uncertainty not modelled by the covariates (Section \ref{sec:components}). The case of known dispersion is treated in Section \ref{sec:known}. Section \ref{sec:dispersion_EDM} presents an elicitation procedure for an unknown dispersion parameter $\phi$ in Eq. \eqref{eq:model} that applies to any exponential dispersion model. Procedure \ref{alg:dispersion} summarises the structured elicitation for the random component. 

\subsection{Known Dispersion}\label{sec:known}

If the dispersion parameter $\phi$ in Eq. \eqref{eq:model} is known, then let $s, r \rightarrow \infty$ in Eq. \eqref{eq:model} such that $s/r = 1/{\phi}$. The marginal distribution of $\beta$ is then multivariate normal by Eq. \eqref{eq:beta_marginal} and $\eta \sim N(m, {\phi}V)$ in Eq. \eqref{eq:eta}. 

\subsection{Unknown Dispersion}\label{sec:dispersion_EDM}

As noted in Section \ref{sec:introduction}, experts should only be asked to elicit observable quantities and should not be asked to predict the variance of a response directly \citep[e.g.,][]{Kadane1998, Garthwaite2000}. The elicitation of a prior for an unknown dispersion parameter $\phi$ is based on the predicted sample mean $\bar{\psi}$ of a hypothetical experiment conditional on the systematic component. This conditioning isolates the observation uncertainty modelled by the random component of the GLM. 

Let the elicitation target follow an exponential dispersion model with variance function $v(\mu)$. If $\psi_j \mid \mu, \phi  \sim ED(\mu, \phi)$ for $j = 1:w$ then the sample mean is closed under convolution \citep{Jorgensen1987},
\begin{equation}
	\bar{\psi} = \frac{1}{w}\sum_{j = 1}^w\psi_j \sim ED\left(\mu, \frac{\phi}{w}\right)\,,\label{eq:EDmean}
\end{equation}
and the central limit theorem applies as the sample size $w \rightarrow \infty$ such that
\begin{equation}
	({\bar{\psi} - \mu})/{\sqrt{\phi/w}} \overset{d}{\rightarrow} N(0, v(\mu))\,.\label{eq:CLT}
\end{equation}
From Eqs. \eqref{eq:model_lambda} and \eqref{eq:CLT}, the sample mean conditional on a realised systematic component $\hat{\mu}_0 \in \Omega$ is for large $w$ approximately univariate generalised t or equivalently t distributed,
\begin{equation}
	\bar{\psi} \mid \hat{\mu}_0, w \sim St_1(\hat{\mu}_0, v(\hat{\mu}_0)/w, r, s) = t_1(\hat{\mu}_0, v_{\phi}(\hat{\mu}_0, w), s)\,, \,
	v_{\phi}(\hat{\mu}_0, w) = rv(\hat{\mu}_0)/(ws)\,,\label{eq:sample_mean}
\end{equation}
with scale $v_{\phi}(\hat{\mu}_0, w)$.
In Eq. \eqref{eq:sample_mean}, the realised systematic component $\hat{\mu}_0$  is the predicted response for an arbitrary scenario, denoted say by $u_0^\top$, in the absence of observation uncertainty. Theoretically, any value $\hat{\mu}_0 \in \Omega$ may be chosen by either the facilitator or the expert. Practically, the expert must be able to hypothesise a scenario of covariates for which the choice of $\hat{\mu}_0$ is plausible. The conditional elicitation of $\bar{\psi}$ given $\hat{\mu}_0$ includes all and only sources of uncertainty for prediction that are unexplained by the covariates.

Conditional on a value $\hat{\mu}_0$ consistent with a hypothetical scenario, the elicitation of the sample mean $\bar{\psi}$ given $w$ independent realisations of the target $\psi$ parametrises Eq. \eqref{eq:model_lambda}. The expert is asked to provide two central credible intervals of probabilities  for the conditional sample mean $\bar{\psi}$ given $\hat{\mu}_0$ and $w$. The two central credible intervals have probabilities $\alpha_1 < \alpha_2$ with corresponding lower bounds $d_1 > d_2$ defined by
\begin{equation*}
	P\left(\bar{\psi} \leq d_k \mid \hat{\mu}_0, w\right) = P\left(\bar{\psi} > 2\hat{\mu}_0 - d_k \mid \hat{\mu}_0, w \right) = \frac{1 - \alpha_k}{2}\,, \quad k \in \{1, 2\}\,, \quad \alpha_1 < \alpha_2\,.\label{eq:varfun}
\end{equation*}
From Eq. \eqref{eq:sample_mean}, the quantile function for the sample mean is
\begin{equation}
	F^{-1}(q \mid \hat{\mu}_0, w, r, s) = \hat{\mu}_0 + \sqrt{\frac{r v(\hat{\mu}_0)}{ws}} T_1^{-1}(q \mid s)\,,\quad q\in(0, 1)\,,
	\label{eq:samplemean_q}
\end{equation}
where $T_1^{-1}(\cdot \mid s)$ is the quantile function of the univariate Student's t distribution with $s$ degrees of freedom, location of zero and scale of one. Given the choices of $d_k$ and $\alpha_k$ for $k \in \{1, 2\}$, the elicited parameter $s$ is obtained as the solution to the relation
\begin{equation}
	\frac{d_1 - \hat{\mu}_0}{d_2 - \hat{\mu}_0} = \frac{F^{-1}\left(\frac{1 - \alpha_1}{2}  \mid \hat{\mu}_0, w, r, s\right) - \hat{\mu}_0}{F^{-1}\left(\frac{1 - \alpha_2}{2}  \mid \hat{\mu}_0, w, r, s\right) - \hat{\mu}_0} = \frac{T_1^{-1}\left(\frac{1 - \alpha_1}{2} \mid s\right)}{T_1^{-1}\left(\frac{1 - \alpha_2}{2} \mid s \right)}  
	< \frac{\Phi^{-1}\left(\frac{1 - \alpha_1}{2} \right)}{\Phi^{-1}\left(\frac{1 - \alpha_2}{2} \right)}\,,\label{eq:s}
\end{equation}
where $\Phi^{-1}(\cdot)$ is the quantile function of the limiting standard normal as $s, r \rightarrow \infty$ in Eq. \eqref{eq:samplemean_q}. The elicited bounds $d_1$ and $d_2$ are held consistent with the generalised t approximation by the upper bound in Eq. \eqref{eq:s}, see also \citet{Kadane1980} for application of this relation to the elicitation of the degrees of freedom parameter in the normal linear model. Given the degrees of freedom $s$ and the elicited quantiles $d_1$ and $d_2$, the rate parameter $r$ is obtained from Eqs. \eqref{eq:sample_mean} and \eqref{eq:samplemean_q} by
\begin{equation}
	v_{\phi}(\hat{\mu}_0, w)  = \left(\frac{d_k - \hat{\mu}_0}{T_1^{-1}([1 - \alpha_k]/2 \mid s)}\right)^2\quad \implies \quad 
	r = \frac{v_{\phi}(\hat{\mu}_0, w)ws}{v(\hat{\mu}_0)}\,, \quad k \in \{1, 2\}\,.\label{eq:r}
\end{equation}
The choices of $\alpha_1$ and $\alpha_2$ should be sufficiently different to elicit both the degrees of freedom $s$ from Eq. \eqref{eq:s}, which  controls the heaviness of the tails, and the scale $v_{\phi}(\hat{\mu}_0, w)$ from Eq. \eqref{eq:r}. 

Several advantages of this elicitation approach for unknown $\phi$ are now summarised: 
\begin{enumerate}
	\itemsep0em 
	\item the approximate density function of the sample mean is unimodal and symmetric, which agrees with general expectations often expressed by experts \citep{Winkler1967}; 
	\item the approximating t distribution can be made arbitrarily close to the distribution of the sample mean under $f$ in Eq. \eqref{eq:model_obs} by choice of $w$ large enough, and the quality of this approximation is assessable;
	\item the completed elicitation can be used to support prior parametrisations for alternative observation models (Section \ref{sec:induced-prior}); and
	\item the same elicitation strategy for unknown dispersion applies to any exponential dispersion model, whether the support is discrete, continuous or even semi-continuous (see Section \ref{sec:CP} for an example of the latter).
\end{enumerate}
Techniques for  evaluating the quality of the t approximation are discussed  below. 

A check on the expert's understanding of the amount of information in a sample (Section \ref{sec:components}) can be assessed by the facilitator through trialling different values of $w$, and confirming that the expert correspondingly adjusts the predictions for the sample mean. The facilitator may also wish to guide choices of finite $w$ and $\hat{\mu}_0$  that are robust to the CLT approximation of Eq. \eqref{eq:CLT}. Based on a facilitator's point estimate for $\phi$, a bound on the approximation error of Eq. \eqref{eq:CLT} is available through the Berry-Esseen inequality,
\begin{equation}
	\sup_z \left|P\left(\frac{\sqrt{w}(\bar{\psi} - \hat{\mu}_0)}{\sigma} \leq z\right) - \Phi(z)\right| \leq \frac{K \breve{\mu}^3}{\sqrt{w}} \leq  \frac{K \sqrt{\breve{\mu}^4}}{\sqrt{w}}\,,
	\label{eq:BE}
\end{equation}
where $\sigma = \sqrt{\phi v(\hat{\mu}_0)}$ and $\breve{\mu}^k = \mathbb{E}\left[\left|\psi - \hat{\mu}_0\right|^k\right]/\sigma^{k}$ for $k \in \{3, 4\}$. A recent estimate is $K = 0.469$  \citep{Shevtsova2017}. The upper bound on the Berry-Esseen inequality in Eq. \eqref{eq:BE} uses the more accessible kurtosis $\breve{\mu}^4$  or excess kurtosis $\breve{\mu}^4 - 3$ (see Section \ref{sec:BEbound} for details). This upper bound may assist the facilitator in suggesting values for $\hat{\mu}_0$ and $w$ that improve the generalised t approximation for the sample mean.

The discrepancy of the sample mean distribution with respect to the approximating distribution may also be considered. Two options are provided that use Monte Carlo samples of $\lambda$ from Eq. \eqref{eq:model_lambda} followed by composition sampling for the sample mean from Eq. \eqref{eq:EDmean}. The latter step is conditional on $\lambda$ and so only the ability to sample from natural exponential families is required. First, the  distribution function of the sample mean ${F}(\bar{\psi} \mid \hat{\mu}_0, w, s, r)$  is estimated with accompanying confidence intervals for the Monte Carlo error (see Section \ref{sec:discrepancy}). The estimated  distribution function of the sample mean, $\hat{F}(\bar{\psi} \mid \hat{\mu}_0, w, s, r)$, and its generalised t approximation are then compared. A numerical measure of discrepancy is provided by the Kolmogorov distance,
\begin{equation}
	\sup_x \left| \hat{F}(x \mid \hat{\mu}_0, w, s, r) - T_1\left(\frac{x - \hat{\mu}_0}{\sqrt{v_{\phi}(\hat{\mu}_0, w)}} \mid s\right) \right|\,.\label{eq:KD}
\end{equation}
Second, the expected logarithm of the posterior odds against the generalised t approximation is estimated (see Section \ref{sec:KLdisp}). Assuming equal prior odds, this measure is equivalent to the Kullback-Leibler divergence \citep{Kullback1951}, or expected log Bayes factor, for the joint distribution of the sample mean and dispersion relative to the approximation from Eq. \eqref{eq:CLT},
\begin{equation}
	D_{\bar{\psi}, \lambda} = 
	\mathbb{E}_{\bar{\psi}, \lambda \mid \hat{\mu}_0, w, s, r}\left[ 
	\log \frac{f(\bar{\psi} \mid \hat{\mu}_0, \frac{1}{w \lambda})G(\lambda \mid \frac{s}{2}, \frac{r}{2})}{N(\bar{\psi} \mid \hat{\mu}_0, \frac{1}{w \lambda})G(\lambda \mid \frac{s}{2}, \frac{r}{2})}\right]\,.
	\label{eq:KLdisp}
\end{equation}
\citet[][Appendix B]{Jeffreys1961} and \citet{Kass1995} suggest thresholds for the acceptance or rejection of hypotheses based on the Bayes factor. Eqs. \eqref{eq:KD} and \eqref{eq:KLdisp} can be assessed before or after an expert has contributed bounds $d_1$ and $d_2$. In the former case, the facilitator may suggest $w$ based on experimentation for a plausible choice of $s$ and $r$. In the latter case, if the elicited discrepancies are unacceptably large then the elicitation can be iteratively reassessed for the unknown dispersion (Procedure \ref{alg:dispersion}). 

\section{Systematic Component}\label{sec:systematic}

The elicitation of the systematic component and associated process uncertainty  conditions on the scenarios documented by $U$ (Section \ref{sec:components}). The $n \times 1$ systematic component $\mu$ is indexed by scenario, such that $\mu_i = \mu \mid u_i^\top$, where $u_i^\top$ is the $i$\textsuperscript{th} row of $U$ that defines the $i$\textsuperscript{th} scenario for $i = 1:n$. Elicitation for the systematic component begins with the marginal distributions of $\mu$ (Section \ref{sec:marginal}). The dependencies among scenarios are then modelled with a canonical vine copula (Section \ref{sec:DCMP}). The scenarios additionally condition on hypothetical values of the systematic component $\hat{\mu}_{1:l}$ realised at the first $l$ scenarios for $1 \leq l < n$ (Section \ref{sec:estimation}). At each successive tree level $\mathcal{T}_l$ in the canonical vine, the expert elicits the conditional distribution of the systematic component $\mu_k \mid \hat{\mu}_{1:l}$ for $k = (l + 1):n$. The canonical vine copula model for dependence is progressively parametrised with conditional median elicitations  such that the expert is never asked to forget hypothetical realised data once admitted into the elicitation session, which accords with the  design principles of the elicitation method (Section \ref{sec:introduction}). The choice of the conditioning values $\hat{\mu}_{1:l}$ for the systematic component is considered in Section \ref{sec:conditioning}. To optionally reduce the number of conditional elicitations, $t$-truncated canonical vine dependence models may be used (Section \ref{sec:truncated-cmps}). A $t$-truncated vine assumes no dependence on hypothetically realised values of the systematic component above tree level $\mathcal{T}_t$. 

\subsection{Elicitation of Marginal Distributions}\label{sec:marginal}

Given the marginal prior of Eq. \eqref{eq:beta_marginal} and the specification $X = I$ in Eq. \eqref{eq:eta}, the linear predictor is distributed as 
\begin{equation}
	\eta \sim St_n(m, V, r, s)\label{eq:eta_prior}
\end{equation}
with $n\times 1$ location vector $m$ and $n\times n$ positive definite scale matrix $V$. The parameters $s$ and $r$ are known from the elicitation of the random component (Section \ref{sec:unknown_dispersion}). Given also the link function $g$ in Eq. \eqref{eq:model_link}, the  density function for the systematic component is
\begin{equation}
	f(\mu \mid m, V, r, s) = St_n\left(g(\mu) \mid m, V, r, s\right)\,|dg(\mu)/d\mu|\,, \quad \mu \in \Omega\,.\label{eq:mu}
\end{equation}
From Eq. \eqref{eq:mu}, expressions for the univariate density, distribution and quantile functions of the systematic component at each scenario in $U$ are available to provide graphical and numerical feedback to the expert, see Eqs. \eqref{eq:mu_density}, \eqref{eq:mu_dist} and \eqref{eq:mu_q} in Section \ref{sec:cond_dists}.

The central credible intervals $(a, b)$ for  $\mu = g^{-1}(\eta)$ defined by
\begin{equation*}
	P\left(\mu_i \leq a_i \right) = P\left(\mu_i > b_i \right) = \frac{1 - \alpha}{2}\,  , \quad (a_i, b_i) \in \Omega\,, \quad i = 1:n\,, \quad 0 < \alpha < 1\,,\label{eq:CI}
\end{equation*}
are elicited, where $P\left(\mu_i \leq z \right) = F_{\mu}(z \mid m_i, V_{ii}, r, s)$ is the marginal distribution function of the systematic component at elicitation scenario $u_i^\top$. The location and scale are
\begin{equation*}
	m_i = \frac{g(a_i) + g(b_i)}{2}\, , \quad V_{ii} = \left(\frac{g(b_i) - m_i}{T_1^{-1}((1 + \alpha)/2 \mid s)}\right)^2\frac{s}{r}\,,
\end{equation*}
because $g$ is invertible. The elicited central credible interval $(a_i, b_i)$ for the systematic component at scenario $u_i^\top$ uniquely parametrises the required marginal distribution of the linear predictor $\eta_i$ for $i = 1: n$. 

The expert iteratively updates each marginal  prior  by adjustment of the central credible interval $(a_i, b_i)$ until the relevant probabilities and quantiles of interest at scenario $u_i^\top$, along with any  graphical feedback, adequately approximate the expert's beliefs given the model assumptions. This elicitation protocol is repeated for each elicitation scenario $u_i^\top$, $i = 1:n$, to induce the $n$ marginal  priors for the underlying linear predictor $\eta$. 
The above elicitation steps are summarised in Procedure \ref{alg:CI}.

\subsection{Dependence and Canonical Vine Copula }\label{sec:DCMP}

The conditional distribution of the  linear predictor $\eta_{l:k} \mid \eta_{1:(l-1)}$ for $1 \leq l \leq k \leq n$, where $\eta_{1:0} = \emptyset$, has the generalised t distribution \citep[][Ch. 5]{Kotz2004}, 
\begin{equation}
	\eta_{l:k} \mid \eta_{1:(l-1)} \sim St_{k - l + 1}\left(m_{l:k \mid 1:(l - 1)}, V_{l:k, l:k \mid 1:(l-1)}, r + \zeta(\eta_{1:(l-1)}), s + l - 1 \right)\label{eq:cond_eta}
\end{equation}
with conditional location vector
\begin{equation}
	m_{l:k \mid 1:(l - 1)} = m_{l:k} + V_{l:k,1:(l - 1)} V_{1:(l - 1), 1:(l - 1)}^{-1}\left(\eta_{1:(l - 1)} - m_{1:(l - 1)}\right)
	\label{eq:cond_m}
\end{equation}
and conditional scale matrix
\begin{equation}
	V_{l:k,l:k \mid 1:(l - 1)} = V_{l:k,l:k} - V_{l:k,1:(l - 1)} V_{1:(l - 1), 1:(l - 1)}^{-1}V_{1:(l - 1), l:k}\,,\label{eq:cond_V}
\end{equation}
where $\zeta(\eta_{1:(l-1)}) = (\eta_{1:(l-1)} - m_{1:(l - 1)})^\top V_{1:(l-1), 1:(l-1)}^{-1}(\eta_{1:(l - 1)} - m_{1:(l - 1)})$. 
Conditional on the index parameter $\lambda$, the linear predictor $\eta$ is multivariate normal with mean $m$ and covariance $V/\lambda$ by Eq. \eqref{eq:eta}. The conditional correlation \citep[][Ch. 2]{Joe2015} for this multivariate normal distribution is 
\begin{align}
	\rho_{l,k|1:(l - 1)} =& \left({V_{k,l} - V_{l,1:(l - 1)}V_{1:(l - 1),1:(l - 1)}^{-1}V_{1:(l - 1),k}}\right)\nonumber\\
	&\times\left({V_{l,l} - V_{l,1:(l - 1)}V_{1:(l - 1),1:(l - 1)}^{-1}V_{1:(l - 1),l}}\right)^{-1/2}\label{eq:cond_corr}\\
	&\times\left({V_{k,k} - V_{k,1:(l - 1)}V_{1:(l - 1),1:(l - 1)}^{-1}V_{1:(l - 1),k}}\right)^{-1/2},\quad 1 \leq l \leq k \leq n\,.\nonumber
\end{align}
For $k\neq l$, Eq. \eqref{eq:cond_corr} is equivalent to the partial correlation of variates $\eta_k$ and $\eta_l$ conditional on $\lambda$ and the realised variates $\eta_{1:(l - 1)}$. If $l = 1$ then Eq. \eqref{eq:cond_corr} reduces to the marginal correlation between $\eta_k$ and $\eta_{l}$, given $\lambda$, that is defined by $\rho_{l,k} = V_{k,l}/\sqrt{V_{l,l}V_{k,k}}$.

\citet{Bedford2002} use a specific selection of $\binom{n}{2}$ partial correlations to develop a regular {vine} copula that defines a positive definite correlation matrix $R$, such that each partial correlation has range $(-1, 1)$. There are various formulations of regular vines that each provide a different parametrisation of an equivalent positive definite correlation matrix $R$ \citep[][Ch. 3]{Joe2015}. On example is the \emph{canonical} vine copula \citep{Bedford2001} that can be represented by an upper triangular matrix of partial correlations,
\begin{equation*}
	\Rho = \begin{bmatrix}
		1 & \rho_{1,2} & \rho_{1,3} & \ldots & \rho_{1,n-1} & \rho_{1, n}\\
		& 1		   & \rho_{2, 3 | 1} & \ldots & \rho_{2, n- 1| 1} & \rho_{2, n |1}\\
		& 		   & 				 & \ddots &  & \vdots \\
		&			&		&	&	1		& \rho_{n - 1, n | 1:(n-2)}		\\
		0		& 		&		&		&			&			1\\
	\end{bmatrix}.
\end{equation*}
Each row  of the matrix $\Rho$ corresponds to a level of dependency, denoted by $\mathcal{T}_l$ and referred to as the tree level \citep[][]{Bedford2001, Bedford2002, Joe2015}. The tree level $\mathcal{T}_l$ is the set of conditional correlations between $\eta_l$ and each member of the set $\eta_{(l + 1):n}$ conditioned on the set $\eta_{1:(l -1)}$. The canonical vine has the same conditioning variables within a tree level, and exactly one variable is added to the conditioning set when moving up one level \citep{Bedford2001}. \citet{Joe2015} provides an algorithm for the mapping $\Rho \rightarrow R$ (see Algorithm \ref{alg:PR} in Section \ref{sec:bounds_median}). 

The canonical vine is particularly advantageous for conditional elicitation. The canonical vine ensures that the expert is never asked to forget hypothetical realised data admitted into the  elicitation session, in accord with the design principles of the elicitation method (Section \ref{sec:introduction}). Regular vines other than the canonical vine adopt a different partial correlation structure $\Rho$ and introduce more than one conditioning variable at lower tree levels \citep[][Ch. 3]{Joe2015}. On the opposite extreme from the canonical vine is the $D$-vine, for example, where the number of unique conditioning variables at each tree level is maximised within the regular vine constraints. For regular vines other than the canonical vine, the expert is asked at some point to forget conditioning values from prior elicitations, only to have these previously considered conditioning values reintroduced at the same tree level or at higher tree levels. This complication is avoided by means of the canonical vine copula.

\subsection{Elicitation of Canonical Vine Copula}\label{sec:estimation}

At tree level $\mathcal{T}_{l}$, let  $c_{k|1:l}$ denote for $k > l$ the conditional median of the systematic component at scenario $u_{k}^\top$ given hypothetical values $\hat{\mu}_{1:l} = g^{-1}(\hat{\eta}_{1:l})$ of the systematic component realised at the first $l$ scenarios. The conditional median is defined by
\begin{equation*}
	P\left(\mu_{k} \leq c_{k \mid 1:l}  \left. \mid \hat{\mu}_{1:l}\right.\right) = 0.5\,, \quad
	1\leq l < k \leq n\, ,  \quad c_{k \mid 1:l} \in \Omega\,.\label{eq:centre}
\end{equation*}
The sequence of elicitations for conditional medians progressively elicits the upper triangular entries of the array 
\begin{equation*}
	C = \begin{pmatrix}
		c_1 & c_{2 \mid 1} & c_{3 \mid 1} & \ldots & c_{n-1 \mid 1} & c_{n \mid 1}\\
		& c_2		   & c_{3 \mid  1:2} & \ldots & c_{n- 1 \mid  1:2} & c_{n \mid 1:2}\\
		& 		   & 				 & \ddots &  & \vdots \\
		&			&		&	&	c_{n-1}		& c_{n \mid  1:(n-1)}		\\
		0		& 		&		&		&			&			c_n\\
	\end{pmatrix}\,.
\end{equation*}
The elicitation sequence begins in the top row for tree level $\mathcal{T}_1$ and continues until the second to bottom row for tree level $\mathcal{T}_{n - 1}$. Within each row (tree level) the conditional elicitations may be completed in any order, but would typically be conducted in sequence. The main diagonal entries of $C$ are designated as the elicitations at tree level $\mathcal{T}_0$, where the marginal medians of the systematic component are  $\diag[C] = g^{-1}(m)$ because $g$ is invertible. The  vector $m$ was previously elicited in Section \ref{sec:marginal}. Moreover, the conditional median of the systematic component is $m_{k|1:l} = g(c_{k \mid 1:l})$. 

Given hypothetically realised values of the systematic component $\hat{\mu}_{1:l} = g^{-1}(\hat{\eta}_{1:l})$, Eq. \eqref{eq:cond_m} with correspondingly adjusted indices shows that the difference between the conditional and marginal medians for $\eta_k$  is
\begin{align}
	m_{k|1:l} - m_k &=  V_{k,1:l} V_{1:l, 1:l}^{-1}\left(\hat{\eta}_{1:l} - m_{1:l}\right)\,, \quad 1 \leq l < k \leq n\,. \label{eq:median_diff}
\end{align}
Let $h_{1:l} = V_{1:l, 1:l}^{-1}\left(\hat{\eta}_{1:l} - m_{1:l}\right)$. Solving for $V_{k, l}$ in Eq. \eqref{eq:median_diff} with realisations for the linear predictor such that $\hat{\eta}_{1:l} \neq m_{1:l}$ then obtains
\begin{equation}
	V_{k, l} = \begin{cases}
		\left(m_{k \mid 1} - m_k\right)/h_1 & \textrm{ if } l = 1\, ,\\
		\left[\left(m_{k \mid 1:l} - m_k\right) - V_{k, 1:(l - 1)}h_{1:(l - 1)}\right]\big\slash h_l & \textrm{ otherwise.}
	\end{cases}
	\label{eq:Vkl}
\end{equation}
At tree level $\mathcal{T}_l$, the entries of $m$, $V$ and $h$ that appear on the right hand side of Eq. \eqref{eq:Vkl} have already been derived from elicitations completed for  the marginal elicitations (Section \ref{sec:marginal}) and  conditional medians elicited in $C$ at lower tree levels $\mathcal{T}_{1:(l - 1)}$. Apart from the conditional median $m_{k|1:l}$ that is the current focus of elicitation, there are no other unknowns on the right hand side of Eq. \eqref{eq:Vkl} at tree level $\mathcal{T}_l$. 

Given $V_{k, l}$, the scale of $\eta_k$ conditional on $\hat{\eta}_{1:l}$ is 
\begin{equation}
	V_{k,k \mid 1:l} = V_{k,k} - V_{k,1:l} V_{1:l, 1:l}^{-1}V_{1:l, k}\,,\label{eq:Vkkl}
\end{equation}
which follows from Eq. \eqref{eq:cond_V} with indices adjusted to account for conditioning on $\hat{\eta}_{1:l}$.
From the identity $V_{k,k \mid 1:l} = V_{k,k \mid 1:(l - 1)}(1 - \rho_{l,k \mid 1:(l - 1)}^2)$ \citep[][Ch. 5]{Whittaker1990} and the previously elicited conditional scale entry $V_{k,k \mid 1:(l - 1)}$ at the preceding tree level $\mathcal{T}_{l - 1}$, the constraint $|\rho_{l,k \mid 1:(l - 1)}| < 1$ (Section \ref{sec:DCMP}) requires 
\begin{equation}
	0 < V_{k,k \mid 1:l} \leq V_{k,k \mid 1:(l - 1)}\,.\label{eq:check}
\end{equation}
The elicited conditional scale for $\eta_{k \mid 1:l}$ at the current tree level $\mathcal{T}_l$ must be positive and  cannot exceed the conditional scale for $\eta_{k \mid 1:(l-1)}$ previously determined at tree level $\mathcal{T}_{l - 1}$. The consistency check of Eq. \eqref{eq:check} is available for feedback to the expert at tree level $\mathcal{T}_l$. Feasible bounds on the conditional median $c_{k|1:l} = g^{-1}(m_{k|1:l})$ may be graphically or numerically presented to the expert at the start of each elicitation (Section \ref{sec:bounds_median}). 

At each tree level $\mathcal{T}_l$, the completed elicitation $c_{k|1:l} = g^{-1}(m_{k \mid 1:l})$  induces the scale entry $V_{k, l}$ from Eq. \eqref{eq:Vkl}, which having passed the consistency check of Eq. \eqref{eq:check} then defines the partial correlation $\Rho_{lk} = \rho_{l, k \mid 1:(l - 1)}$ by Eq. \eqref{eq:cond_corr}. Thus the successive elicitation of each row of entries along the upper triangular part of $C$ completes the corresponding row in $\Rho$. The univariate conditional density, distribution and quantile functions of the systematic component are also available to provide graphical and numerical feedback to the expert, see Eqs. \eqref{eq:mucond_density}, \eqref{eq:mucond_dist} and \eqref{eq:mucond_q} in Section \ref{sec:cond_dists}. The conditional elicitation steps are summarised in Procedure \ref{alg:cond}. The choice of conditioning hypothetical realisations $\hat{\mu}_{1:l}$ is discussed in Section \ref{sec:conditioning}.

\subsection{Conditioning Values for Systematic Component}\label{sec:conditioning}

In Section \ref{sec:estimation}, the conditioning values of the systematic component  by Eq. \eqref{eq:Vkl} need only be specified such that $g(\hat{\mu}_j) = \hat{\eta}_j \neq m_j$ with $\hat{\mu}_j \in \Omega$ for $j = 1:l$. From the expert's perspective, however, it is important to choose conditioning values for the target that are not overly surprising or extreme so as to avoid severe or uneven constraints on the conditional scale parameters elicited at different tree levels by Eq. \eqref{eq:check}. 

One choice for the conditioning values of the systematic component is as follows. At the start of elicitations for tree level $\mathcal{T}_l$, set $\hat{\mu}_{l}$ equal to one of the bounds of the conditional central credible interval $(a_{l \mid 1:(l - 1)}, b_{l \mid 1:(l - 1)})$ of probability $\alpha$ defined by the elicited distribution of the systematic component from the previous tree level $\mathcal{T}_{l - 1}$.  Choosing the same probability level $\alpha$ as in the marginal elicitations (Section \ref{sec:marginal}) efficiently builds on the expert's earlier experience at previous tree levels. The conditioning values reflect the non-increasing, and typically decreasing, uncertainty in the systematic component  with increasing tree level by Eq. \eqref{eq:check}. 
The selection of the upper versus lower bound from the conditional central credible interval for the hypothetical conditioning variable  may at each tree level be  chosen by the facilitator or the expert, or selected systematically or randomly. 

The above approach avoids outliers and is always applicable, and so it is a reasonable default method for the choice of conditioning values. But the common choice of $\alpha$ in the above approach is not the only possibility. An alternative choice for the realised values is demonstrated in the elicitation case study of Section \ref{sec:study} given unknown dispersion for a bounded elicitation target. In that example, the conditioning values $\hat{\mu}_{1:l}$ are selected from the bounds of conditional credible intervals of probability $\alpha$ induced by Eq. \eqref{eq:eta} with fixed dispersion of one. The magnitude of the differences between the conditioning values and elicited medians are then determined by the elicited scale $V$ for the linear predictor, which screens out noise introduced by the unknown dispersion. 

The elicitation for the canonical vine copulas sequentially builds on conditioning values, and  thereby avoids scenarios where an expert is asked at some point to forget conditioning values from prior elicitations (Section \ref{sec:introduction}). This sequential aspect means that the ordering of scenarios can also be an important consideration for an elicitation session. The ordering of conditional elicitations is discussed further in the context of truncated vine dependence models (Section \ref{sec:truncated-cmps}).

\subsection{Truncated Canonical Vine Dependence Model}\label{sec:truncated-cmps}

The $t$-truncated canonical vine replaces dependencies at tree levels greater than $t$ with conditional independence assumptions \citep{Joe2015}. Truncation after tree level $\mathcal{T}_t$ sets conditional correlations to zero in upper tree levels $\mathcal{T}_{(t + 1):(n - 1)}$. The truncated partial correlation matrix $\Rho(t)$ has entries
\begin{equation*}
	\Rho_{i,j}(t) = \begin{cases}
		0 & \textrm{if } i > t \textrm{ and }  j > i\\
		\Rho_{i,j} & \textrm{otherwise}
	\end{cases}\,, \quad  i, j = 1, \ldots n\,, \quad t \in \left\{0, \ldots, n - 1\right\}\,.
\end{equation*}
Application of Algorithm \ref{alg:PR} to $\Rho(t)$ returns the corresponding truncated correlation matrix $R(t)$. The truncated elicited scale matrix of Eq. \eqref{eq:eta} is then
\begin{equation*}
	V(t) = \diag[V]^{1/2}R(t)\diag[V]^{1/2}\,.
\end{equation*}
The diagonal entries of $V(t)$ were previously elicited by the marginal elicitations of Section \ref{sec:marginal} such that $\diag[V(t)] = \diag[V]$ for any choice of truncation. Procedure \ref{alg:cond} allows for truncation  during the elicitation session.  

Truncation after level $\mathcal{T}_0$ means that the elicitation for the systematic component need not progress beyond Section \ref{sec:marginal}. If the dispersion parameter is known then the induced prior is an ICMP (Section \ref{sec:glm-icmp}). Truncation after level $\mathcal{T}_{n- 1}$ corresponds to no truncation, that is, $\Rho = \Rho(n - 1)$, $R = R(n - 1)$ and $V = V(n - 1)$. A truncated vine ignores dependencies that would be elicited by considering additional realisations of the systematic component at higher levels. 

\citet{Kurowicka2011} suggests placing high absolute (partial) correlations in low tree levels so that smaller absolute correlations occur in the higher tree levels that are truncated. The sequence of introduced conditioning values during the elicitation can therefore be important. Those scenarios expected to have high dependency with other scenarios should have conditioning values introduced at lower tree levels early in the elicitation session to minimise the loss of information induced by a truncated dependence model. Depending on time and resources the elicitation session may be stopped before all tree levels are completed, and the truncated dependence model then invoked. Nevertheless, it should be noted that dependency information may yet be present at higher tree levels, which would not be revealed without conducting the corresponding conditional elicitations. Truncated vine dependence models are evaluated in the elicitation case study (Section \ref{sec:study}).

\section{Induced Prior}\label{sec:induced-prior}

At the conclusion of the elicitation session, the facilitator has elicited the set of parameters $\theta = \left\{s, r, m, V\right\}$ (Section \ref{sec:specs} and  Procedure \ref{protocol}). The parameters $s$ and $r$ are elicited in Section \ref{sec:unknown_dispersion}. The location $m$ and scale $V$ are elicited  in Section \ref{sec:systematic} for the systematic component $\mu$ conditional on $s$, $r$, link function $g$ and elicitation scenarios $U$. The elicited scale matrix $V$ may be truncated at level $t \in \{0, \ldots, n - 1\}$ in Section \ref{sec:truncated-cmps}; formally, set $V = V(t)$ in the set $\theta$. Let $H$ denote the  model described by Eqs. \eqref{eq:model}  and \eqref{eq:eta} with elicited parameters $\theta$. 

Section \ref{sec:specs} allows for alternative choices of $n \times p$ model matrix $X \in \mathcal{X}$ of full column rank or exponential dispersion model $f' \in \mathcal{F}$. Let $H'$ denote an alternative model described by the choice of exponential dispersion model $f'$ with variance function $v'(\mu)$ in Eq. \eqref{eq:model_obs}, $X \in \mathcal{X}$ in Eq. \eqref{eq:model_link}, $\delta'$ and $\Sigma'$ in Eq. \eqref{eq:model_beta}, and $s'$ and $r'$ in Eq. \eqref{eq:model_lambda}. If the dispersion is known, with value $\phi = r/s$ under $H$ in Section \ref{sec:known}, let $\phi' = r'/s'$ denote its value under $H'$. If the dispersion is unknown, then consistency with the elicited scale $v_{\phi}(\hat{\mu}_0, w)$ of the sample mean distribution conditional on $\hat{\mu}_0$  and $w$ in Section \ref{sec:dispersion_EDM} requires
\begin{equation}
	v_{\phi}(\hat{\mu}_0, w) = \frac{rv(\hat{\mu}_0)}{ws} = \frac{r'v'(\hat{\mu}_0)}{ws'} \, . \label{eq:r_constraint}
\end{equation}
The parameter set of the alternative model $H'$ is $\theta' = \{s', r', \delta', \Sigma'\}$.

The estimation procedure minimises the the Kullback-Leibler divergence or mean information between $H$ and $H'$ per observation from $H$ for $\mu$ and $\lambda$,
\begin{align}
	D(H:H') =& \int \log \frac{f\left(\mu \mid \lambda, \theta, A\right)G\left(\lambda \mid \frac{s}{2}, \frac{r}{2}\right)}{f\left(\mu \mid \lambda, \theta', X\right)G\left(\lambda \mid \frac{s'}{2}, \frac{r'}{2}\right)} f\left(\mu \mid \lambda, \theta, A\right)G\left(\lambda \mid \frac{s}{2}, \frac{r}{2}\right) d\mu d\lambda \nonumber\\
	=& 
	\int \log \frac{N\left(\eta \mid A m,\frac{r}{s\lambda}AVA^\top\right)G\left(\lambda \mid \frac{s}{2}, \frac{s}{2}\right)}{N\left(\eta \mid X\delta', \frac{r'}{s'\lambda}X\Sigma'X^\top \right)G\left(\lambda \mid \frac{s'}{2}, \frac{s'}{2}\right)}\label{eq:KLparam}\\ &\times N\left(\eta \mid Am, \frac{r}{s\lambda}AVA^\top\right)G\left(\lambda \mid \frac{s}{2}, \frac{s}{2}\right) d\eta d\lambda\,,\nonumber
\end{align}
where $A$ is a projection matrix from $H$ onto the column space $\mathbf{C}(X)$ of $X$ such that $P\left(g(\mu) \in \mathbf{C}(X)\right) = 1$. The last equality follows from the invariance property of information \citep[][Ch. 2]{Kullback1959} given  the nonsingular transformation $g: \mu \rightarrow \eta$. The minimisation of Eq. \eqref{eq:KLparam} is with respect to $\theta'$ and the projection matrix $A$. The divergence $D\left(H : H'\right)$ is minimised  by the parameter set $\left\{\hat{\theta}, \hat{A}\right\} = \mathop{\mathrm{arg\,min}}_{\left\{\theta', A\right\}} D\left(H : H'\right)$ (see Section \ref{sec:KL} for details), where $\hat{\theta} = \left\{\hat{s}, \hat{r}, \hat{\delta}, \hat{\Sigma}\right\}$ and
\begin{equation}
	\begin{gathered}
		\hat{\delta} = \left(X^\top V^{-1} X\right)^{-1}X^\top V^{-1} m\,, \quad  \quad \hat{\Sigma} = q^{-1}\left(X^\top V^{-1} X\right)^{-1}\,,\\
		\hat{A} = X\left(X^\top V^{-1} X\right)^{-1}X^\top V^{-1}\,, \quad \hat{s} = s\,,\quad \hat{r} = rq\,,\label{eq:parms}\\
		q = 
		\begin{cases}
			{\phi'}/{{\phi}} & \textrm{if dispersion is known}\\
			{v(\hat{\mu}_0)}/{v'(\hat{\mu}_0)} & \textrm{if dispersion is unknown.}
		\end{cases}
	\end{gathered}
\end{equation} 
If the model matrix $X$ is non-singular then $\hat{A} = I$ such that no information elicited from the expert is lost as the prior is constructed for the unknown coefficients (see Section \ref{sec:KL}). An optimal projection $\hat{A}$ is induced otherwise.

\section{Extensions}\label{sec:extensions}

The generality of statistical models supported by  the  elicitation method is illustrated with examples for  multivariate responses (Section \ref{sec:multivariate}), semi-continuous GLMs with point mass at zero (Section \ref{sec:CP}), extended GLM on the unit interval (Section \ref{sec:DM}),  and overdispersed and zero-inflated count data (Section \ref{sec:over_Poisson}).

\subsection{Multivariate Modelling}\label{sec:multivariate}

Consider $J$ multivariate observations $y_j = [y_{j,1}, \ldots, y_{j,d + 1}]$ of $d + 1$ categories, where for $j = 1:J$ the $k$\textsuperscript{th} entry of  $y_j$ take the value one for category $k$ with probability $\mathbb{E}[y_{j, k}] = p_{j, k} > 0$ and is zero otherwise. The probabilities obey  the constraints $\sum_{k = 1}^{d}p_{j, k} < 1$ and $p_{j, d + 1} = 1 - \sum_{k = 1}^{d}p_{j, k}$. If the $d + 1$ categories are unordered then the observations follow a multinomial distribution, which is an exponential dispersion model with known sample size $\lambda$ \citep{Jorgensen1987}.  Elicitation for multinomial responses is of interest \citep{Elfadaly2017, Wilson2018}. \citet{Elfadaly2020}, for example, adapt the technique of \citet{Kadane1980} to elicit an additive logistic normal prior for the unknown $\beta$ of a multinomial GLM.  The  procedure proposed by \citet{Elfadaly2020} elicits the odds of success $p_{j, k}/p_{j, d + 1}$ for each category $k \in \{1:d\}$ against a reference category $d + 1$. Specification of a multivariate normal distribution on the log odds induces a logistic normal distribution \citep{Aitchison1980} on the $d$ dimensional positive simplex of $p_{j, 1:d}$. 

The statistical elicitation method presented here applies to the logistic normal prior, where the scenarios accommodate the multivariate GLM structure. Identify $d$ rows of the scenario matrix $U$ and the corresponding design points of $X$ with each of the $J$ observations. Let $u(j, k)_i^\top$ and $x(j, k)_i^\top$ denote dependence of scenario $i \in \{1:n\}$ on category $k \in \{1:d\}$ and the set of covariates associated with observation $y_j$. The linear predictor of the multinomial GLM is then $\log(p_{j, k}/p_{j, d + 1})  = x(j, k)_i^\top\beta = \eta_i$. \citet[][Ch. 3]{Fahrmeir1994} describe possible design matrices for multinomial GLMs, see also \citet{Elfadaly2020}. Define the elicitation target $\psi(j,k) = p_{j, k}/p_{j, d + 1}$ for scenario  $u(j, k)_i^\top$. Given logarithmic link function $g$ and dispersion ${\phi} = 1$, Sections \ref{sec:known} and \ref{sec:systematic} then provide the elicitation procedures for an additive logistic normal prior. 

The elicitation method presented here avoids asking experts to provide estimates of conditional central credible intervals (Section \ref{sec:estimation}), as specified in the design principles of Section \ref{sec:introduction}. Also, the method allows reduction of the number of elicited quantities via truncated canonical vine dependence models (Section \ref{sec:truncated-cmps}). A logistic normal prior for the multinomial distribution without covariates is obtained as a special case, where the model matrix is the $d \times d$ identity matrix $X = I$. Other choices for the logistic transformation can be accommodated. For example, the choice $\psi(j,k) = p_{j, k}/(1 - \sum_{k' = 1}^k p_{j, k'})$  with log link $g$ corresponds to a multiplicative logistic normal distribution \citep{Aitchison1982}. See \citet[][Ch. 12]{OHagan2004} for characterisation of multivariate normal priors given log contrasts applied to the multinomial distribution.

If the $d + 1$ categories are ordered then the data are ordinal. The sequential model or continuation ratio model for ordinal data \citep{Fahrmeir1994} is defined by $P(y_j = k \mid y_j \geq k, x(j, k)_i^\top) = h(\eta_i)$, where $h$ is a binary response function and $y_j = k$ if $y_{j, k} = 1$. The sequential model is equivalent to a univariate GLM with unconstrained $\beta$ and known dispersion, where the likelihood is formed from conditional binary responses defined by the $d + 1$ categories \citep[e.g.,][Ch. 9]{Tutz2012}. The procedures of Sections \ref{sec:known} and \ref{sec:systematic} therefore apply. For example, the sequential logit model \citep{Fahrmeir1994} is defined by $\log [P(y_{j} = k \mid x(j, k)_i^\top)/P(y_{j} > k \mid x(j, k)_i^\top)] = \eta_i$. The elicitation method progresses for target $\psi(j,k) = p_{j, k}/(1 - \sum_{k' = 1}^k p_{j, k'})$  with log link $g$ and known dispersion.

\subsection{Power Parameter of Compound Poisson}\label{sec:CP}

Given power parameter $p \in (1, 2)$, the compound Poisson distribution is an exponential dispersion model and Tweedie model with point mass at zero \citep[][Ch. 4; Section \ref{sec:glm}]{Jorgensen1997}. In applications, an unknown power parameter $p$ can be strongly correlated with an unknown dispersion parameter $\phi$ for the compound Poisson family \citep{Peters2009}. For application within the GLM framework, one strategy is to set the  value  of the power parameter a priori \citep{Peel2013}. Consider the case where the power parameter is fixed a priori to a value that is elicited from the expert together with the prior of Eq. \eqref{eq:model_lambda} for the unknown dispersion. 

The elicitation begins with the unknown dispersion $\phi = 1/\lambda$ for the class $\mathcal{F}$ of compound Poisson distributions. Conditional on the choice of systematic component $\hat{\mu}_0$ and sample size $w$, the elicitation for the sample mean $\bar{\psi}$  (Section \ref{sec:dispersion_EDM}) parametrises the shape $s$ and scale $v_\phi(\hat{\mu}_0, w)$ of the approximating univariate t distribution  by Eqs. \eqref{eq:model_lambda}, \eqref{eq:s} and \eqref{eq:r}. However, in Eq. \eqref{eq:r} the parameter $r = {v_{\phi}(\hat{\mu}_0, w)ws}/{v(\hat{\mu}_0)}$ is unknown because it depends on the power parameter $p$ through the variance function $v(\hat{\mu}_0) = \hat{\mu}_0^p$ of the compound Poisson. 
Another elicitation step is therefore required. 

Conditional on $\lambda$ and $\hat{\mu}_0$, the probability that a single response equals zero is $P(\psi = 0 \mid \hat{\mu}_0, \lambda, p) = \exp[-\lambda\hat{\mu}_0^{2 - p}/(2 - p)]$ \citep[see][]{Dunn2008}. Given $\hat{\mu}_0$, let $q_0 = P(\psi = 0 \mid \hat{\mu}_0, p)$ denote the probability that the response is zero by marginalising over the unknown $\lambda$. From Eqs. \eqref{eq:model_lambda} and \eqref{eq:r} with $\hat{\mu}_0$, $w$, $s$ and $v_\phi(\hat{\mu}_0, w)$ obtained in Section \ref{sec:dispersion_EDM}, the probability of zero response  has the unit gamma density function \citep[][]{Grassia1977, Ratnaparkhl1990},
\begin{equation}
	UG(q_0 \mid s, r_p) = \frac{r_p^{s/2}q_0^{r_p -1}(-\log q_0)^{s/2 - 1}}{\Gamma(s/2)}\,, \quad 0 < q_0 < 1\,,\label{eq:UG}
\end{equation}
where $r_p = (2 - p)v_\phi(\hat{\mu}_0, w)ws/(2\hat{\mu}_0^2) > 0$. Conditional on $\hat{\mu}_0$ and the previously elicited $s$ and $v_\phi(\hat{\mu}_0, w)$, an elicited median $c_0$ for the probability of zero response $q_0  = P(\psi = 0 \mid \hat{\mu}_0, s, r_p) = P(\psi = 0 \mid \hat{\mu}_0, s, r, p)$ then parametrises $r_p$ such that
\begin{equation}
	p = 2\left(1 - \frac{r_p\hat{\mu}_0^2}{v_\phi(\hat{\mu}_0, w) w s}\right)\,,\quad
	r = \frac{2r_p\hat{\mu}_0^{2 - p}}{2 - p}\,,\quad
	0 < r_p < \frac{v_\phi(\hat{\mu}_0, w)ws}{2\hat{\mu}_0^2}\,.\label{eq:UGparams}
\end{equation}
The elicited $r_p$ is a solution to $F(c_0 \mid s, r_p) = 1/2$, where $F(x \mid s, r_p) = \int_0^{x} UG(z \mid s, r_p) dz$. This solution is feasible if the upper bound on $r_p$ in Eq. \eqref{eq:UGparams} holds so that $1 < p < 2$, as required for the compound Poisson. Graphical and numerical feedback for $q_0$ based on the unit gamma distribution and density functions, given the elicited  $s$ and $r_p$ of Eq. \eqref{eq:UG}, is available to the expert. The elicitation steps for the power parameter of the compound Poisson distribution  are summarised in Procedure \ref{alg:CP}. Conditional on the elicited $p \in (1, 2)$, elicitation of the systematic component of a compound Poisson GLM then proceeds as in Section \ref{sec:systematic}. Elicitation  of the power parameter of the compound Poisson supports zero-inflated discrete data models in Section \ref{sec:over_Poisson}.

\subsection{Extended Generalised Linear Models}\label{sec:DM}

Extended generalised linear models \citep{Sweeting1981, Jorgensen1983, Jorgensen1997} have an observation model that generalises Eq. \eqref{eq:EDM} to the density
\begin{equation*}
	f(\psi \mid \mu, \phi) = a(\psi, \phi)\exp\left[-d(\psi,\mu)/(2\phi)\right]\,,\quad a(\psi, \phi) > 0\,,
\end{equation*}
with respect to Lebesgue measure for a suitable function $d(\cdot, \cdot)$. This is a dispersion model denoted by $\psi \sim DM(\mu, \phi)$.  As $\phi \rightarrow 0$ in a dispersion model then $DM(\mu, \phi) \approx N(\mu, \phi v(\mu))$ so that $\mu$ is interpreted as a position parameter. Elicitation for the systematic component given $\lambda$ proceeds according to Section \ref{sec:systematic}.
But the convolution formula of Eq. \eqref{eq:EDmean} does not apply to dispersion models in general, and $\mu$ is not necessarily equal to the mean. The procedure of Section \ref{sec:dispersion_EDM} that considers the approximate distribution of the sample mean $\bar{\psi}$ for elicitation of an unknown dispersion parameter in exponential dispersion models therefore does not in general apply to extended GLMs. 

The elicitation procedure for an unknown dispersion in GLMs (Section \ref{sec:dispersion_EDM}), however, does apply to special cases of extended GLMs. One example is the standard simplex distribution on the unit interval \citep{Jorgensen1997a,Jorgensen1997}, which is a dispersion model denoted by $\psi \sim S(\mu, \lambda)$ with
\begin{equation*}
	\mu = \mathbb{E}[\psi]\,,\quad
	d(\psi, \mu) = \frac{(\psi - \mu)^2}{\psi(1 - \psi)\mu^2(1 - \mu)^2}\,,\quad
	v(\mu) = \mu^3(1 - \mu)^3\,, \quad \lambda \in \mathbb{R}^+.
\end{equation*}
The simplex model is of interest because  no exponential dispersion model has bounded support with index set $\lambda \in \Lambda = \mathbb{R}^+$ \citep{Jorgensen1997}. Also, for ${\mu}$ small or large, the variance is approximately $\mathbb{V}[\psi \mid {\mu}, \lambda] \approx v({\mu})/\lambda$  (see Section \ref{sec:svar}) such that  Eq. \eqref{eq:CLT} applies. Given $f$ as a simplex model in Eq. \eqref{eq:model_obs}, prior elicitation for the index parameter $\lambda$ proceeds as in Section \ref{sec:dispersion_EDM}  for the sample mean $\bar{\psi}$ with sample size $w$ conditional on small $\hat{\mu}_0$ or $1 - \hat{\mu}_0$. This strategy is applied to an elicitation target embedded within an overdispersed binomial model (Section \ref{sec:over_Poisson}) by the case study of Section \ref{sec:study}.

\subsection{Overdispersion and Zero-Inflation for Discrete Data}\label{sec:over_Poisson}

Overdispersed discrete data are accommodated, given $\mu$ and $\phi$ in Eq. \eqref{eq:model}, as follows. The data $y$ conditional on the elicitation target $\psi$ are assumed Poisson. The unknown dispersion parameter $\phi$ for the target sets the overdispersion of the count data. Examples of overdispersed Poisson models include \citep{Winkelmann2008}: negative binomial with $y \mid \psi \sim Poisson(\psi)$ and $\psi \sim Gamma(\mu, \phi)$; Poisson-Inverse Gaussian with $y \mid \psi \sim Poisson(\psi)$ and $\psi \sim InverseGaussian(\mu, \phi)$; and Poisson-Lognormal with $y \mid \psi \sim Poisson(e^{\psi})$ and $\psi \sim N(\mu, \phi)$. Elicitation for the target $\psi$ proceeds in the first two cases according to Section \ref{sec:unknown_dispersion} for the dispersion parameter $\phi = 1/\lambda$ and Section \ref{sec:systematic} for the systematic component. These are  examples of Poisson mixture models that use the Tweedie class as a mixing distribution with power parameter $p = 2$ for the negative binomial and $p = 3$ for the Poisson-Inverse Gaussian distribution \citep[][Ch. 4]{Jorgensen1997}. For these first two cases, the elicitation target $\psi$ is the conditional mean of a Poisson model such that  $\mathbb{E}[y | \psi] = \psi$, whereas the third case is instead $\mathbb{E}[y | \psi] = \exp[\psi]$. The Poisson-lognormal is appropriate if the base $B$ logarithm of the conditional mean $\psi_B = \log_B \mathbb{E}[y | \psi]$ is interpretable as an elicitation target. The procedures of Sections \ref{sec:unknown_dispersion} and \ref{sec:systematic} then apply, see Section \ref{sec:lognormal} for details. 

Zero-inflation is another important category for discrete count data \citep{Winkelmann2008}. \citet[][Ch. 4]{Jorgensen1997} suggests the compound Poisson-Poisson mixture that features both overdispersion and zero-inflation due to the zero point mass in the mixing distribution. This model assumes $y \mid \psi \sim Poisson(\psi)$ with conditional mean $\mathbb{E}[y \mid \psi] = \psi$ and compound Poisson mixing distribution with power parameter $1 < p < 2$ for $\psi$. The dispersion and power parameters for the target $\psi$ that set the overdispersion and zero-inflation of $y$ are elicited according to Sections \ref{sec:dispersion_EDM} and \ref{sec:CP}. 

The simplex-binomial mixture \citep[][Ch. 5]{Jorgensen1997} for overdispersed counts assumes $y \sim \linebreak Binomial(k, \psi)$, where $y$ is the number of successes out of $k$ independent trials and $\mathbb{E}[y \mid \psi] = k\psi$, with $\psi \sim S(\mu, \lambda)$ a standard simplex prior on the success probability. The elicitation target $\psi$ follows the standard simplex distribution with dispersion $\phi = 1/\lambda$ (Section \ref{sec:DM}). Overdispersion in $y$ is determined by $\lambda$. The simplex-binomial model with unknown overdispersion is applied in the case study of Section \ref{sec:study}.

\section{Elicitation Case Study}\label{sec:study}

The seagrass species \textit{Zostera nigricaulis} provides an important ecological habitat in Port Philip Bay, which is a large coastal embayment located within an urbanised region of southeastern Australia. Dissolved inorganic nitrogen ($DIN$, mg per L) and total suspended solids ($TSS$, mg per L) are two factors thought to influence the productivity of \textit{Z. nigricaulis} in Port Philip Bay \citep{Hirst2017, Nayar2018}. At low $DIN$ levels \textit{Z. nigricaulis} may exhibit low growth due to lack of nutrients, whereas increased epiphytes load at high $DIN$ levels may suppress seagrass growth by shading and competition for nutrients. Total suspended solids may suppress seagrass  by shading. 

In this case study, a regional model for overdispersed binomial counts $y_i$ of annual mean seagrass percent cover in Port Philip Bay was elicited. The model was defined by
\begin{gather*}
	p(y \mid \psi) = \binom{N}{y}\psi^y(1 - \psi)^{N - y}\,, \quad y \in \{0:N\}\,,\nonumber \\
	\psi \mid \mu, \lambda \sim S(\mu, \lambda)\,, \quad \log \frac{\mu}{1 - \mu} = \eta =  X\beta\,, \quad \beta \mid \lambda \sim N(\delta, \Sigma/\lambda)\,, \quad \lambda \sim G(s/2, r/2)\,,\label{eq:seagrass}
\end{gather*}
with prior structure for $\psi$ according to Eq. \eqref{eq:model}. The elicitation target $\psi$ follows the standard simplex distribution (Section \ref{sec:DM}) such that $f(\psi \mid \mu, 1/\lambda) = S(\psi \mid \mu, \lambda)$ in Eq. \eqref{eq:model_obs}. The  target $\psi$ is the annual mean probability of intersection with seagrass for each of $N = 49$ points in a 0.5 m\textsuperscript{2} quadrat. The annual average percent cover of seagrass is $100\times\psi$. At the observation level, the model is overdispersed binomial conditional on $\mu$ (Section \ref{sec:over_Poisson}).  The linear predictor allows for interactions and quadratic responses with the base 10 logarithm of the annual average $DIN$ in the bottom 1 metre of the water column, and the annual average $TSS$ in the water column (Section \ref{sec:seagrass}). The elicitation scenarios were defined by $DIN$ and $TSS$ (Table \ref{tab:cond}) with corresponding design points that were chosen in consultation with the expert. 

\begin{figure}[t]
	\centering
	\includegraphics[width=0.75\textwidth]{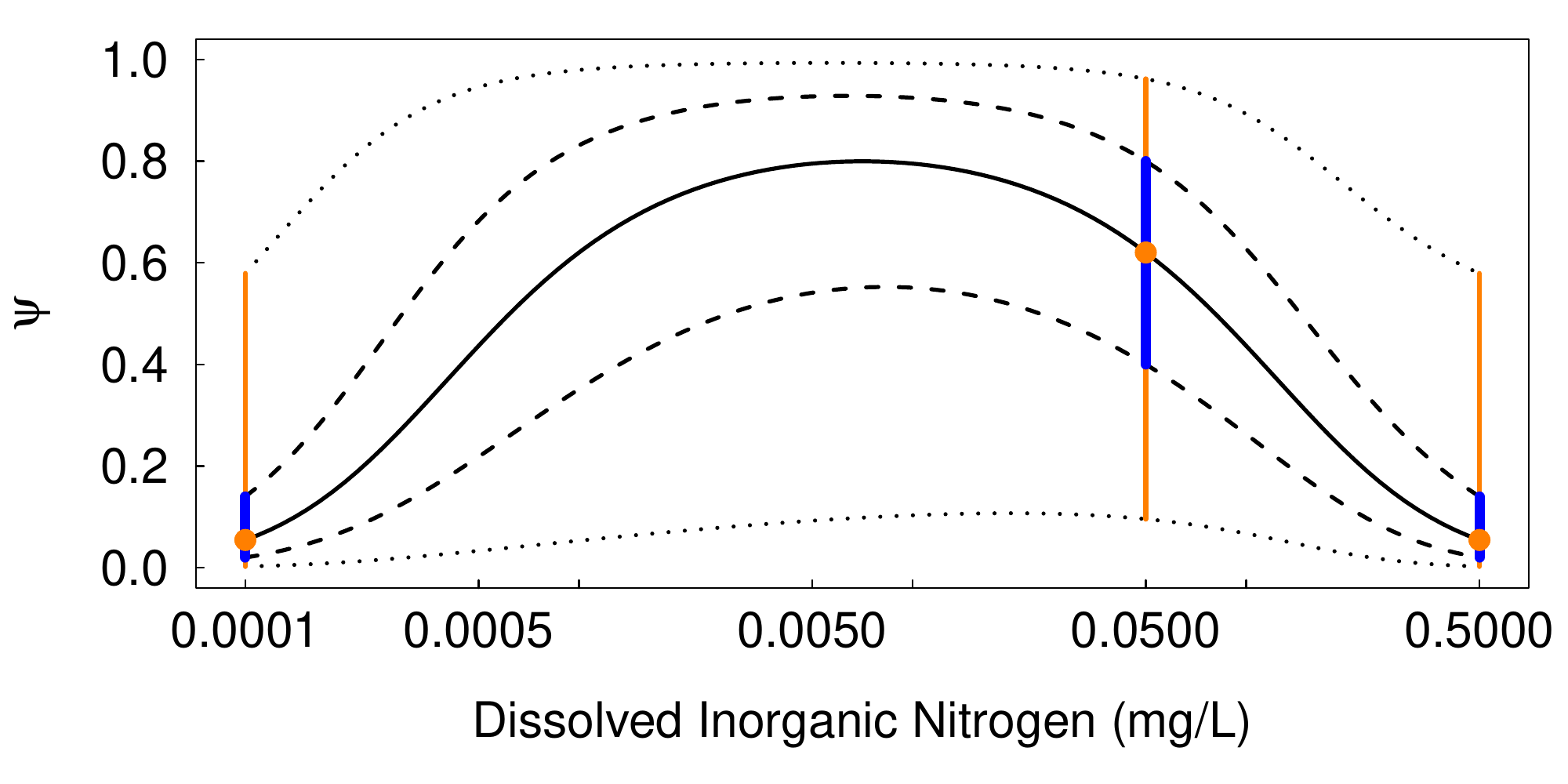}
	\caption{Predicted annual mean probability of seagrass ($\psi$) given $TSS$ equal to 0.10 mg/L. The elicited marginal central credible intervals of probability 1/3 from three scenarios are shown by the three blue vertical lines. The induced marginal medians and central credible intervals of probability 0.8 provided as feedback at these three scenarios are given by the orange points and vertical lines. The predicted median and central credible intervals of probabilities 1/3 and 0.8 across the range of $DIN$ are given by the black solid, dashed and dotted lines, respectively.}\label{fig:pc}
\end{figure}

Probabilistic terms and definitions were reviewed with the expert before the elicitation session (Section \ref{sec:specs}). For the random component, the expert was provided graphical feedback of the approximating Student's t quantiles and probability density function according to Section \ref{sec:dispersion_EDM}. Central credible intervals of probability $1/3$ and $0.90$ were elicited for the sample mean percent cover. Suggested values of the systematic component $\hat{\mu}_0$ emphasised low percent covers between 1\% and 10\% and sample sizes for $w$ between 10 and 800 quadrats (Sections \ref{sec:seagrassRandom}). The discrepancies of the approximation for the final two scenarios considered are plotted in Figure \ref{fig:discrepancy}, where the estimated Kolmogorov distances were $0.035$ and $0.054$ for $\hat{\mu}_0$ equivalent to 1\% and 10\%, respectively. The lower discrepancy for the lower value of $\hat{\mu}_0$ reflects the improved approximation of the generalised t distribution for small $\hat{\mu}_0$ (see Sections \ref{sec:DM}  and \ref{sec:svar}). The random component accounted for variability introduced by covariates not associated with $DIN$ or $TSS$, such as independent factors related to localised disturbance events and the patchy distribution of the target seagrass species in the study region.

For the systematic component, the elicited central credible interval probability level was set to $1/3$ such that the probabilities of the systematic component being above, below or within the interval were equal (Section \ref{sec:seagrassSystematic}). Graphical and numerical feedback for the elicitation target of seagrass percent cover was provided for the $1/3$ and $0.8$ central credible intervals, median, feasible ranges of conditional medians, and plotted density function. The conditional medians were assessed in terms of relative change with respect to previously elicited medians, given newly introduced realised values of the systematic component at successive tree levels. The prior parametrisation for the unknown $\beta$ induced a concave relationship between $DIN$ and seagrass percent cover (Figure \ref{fig:pc}), which is consistent with low percent cover of seagrass at the extremal levels of $DIN$ and high percent cover at intermediate levels of $DIN$ given low levels of $TSS$.

\begin{figure}[t]
	\centering
	\includegraphics[width=0.75\textwidth]{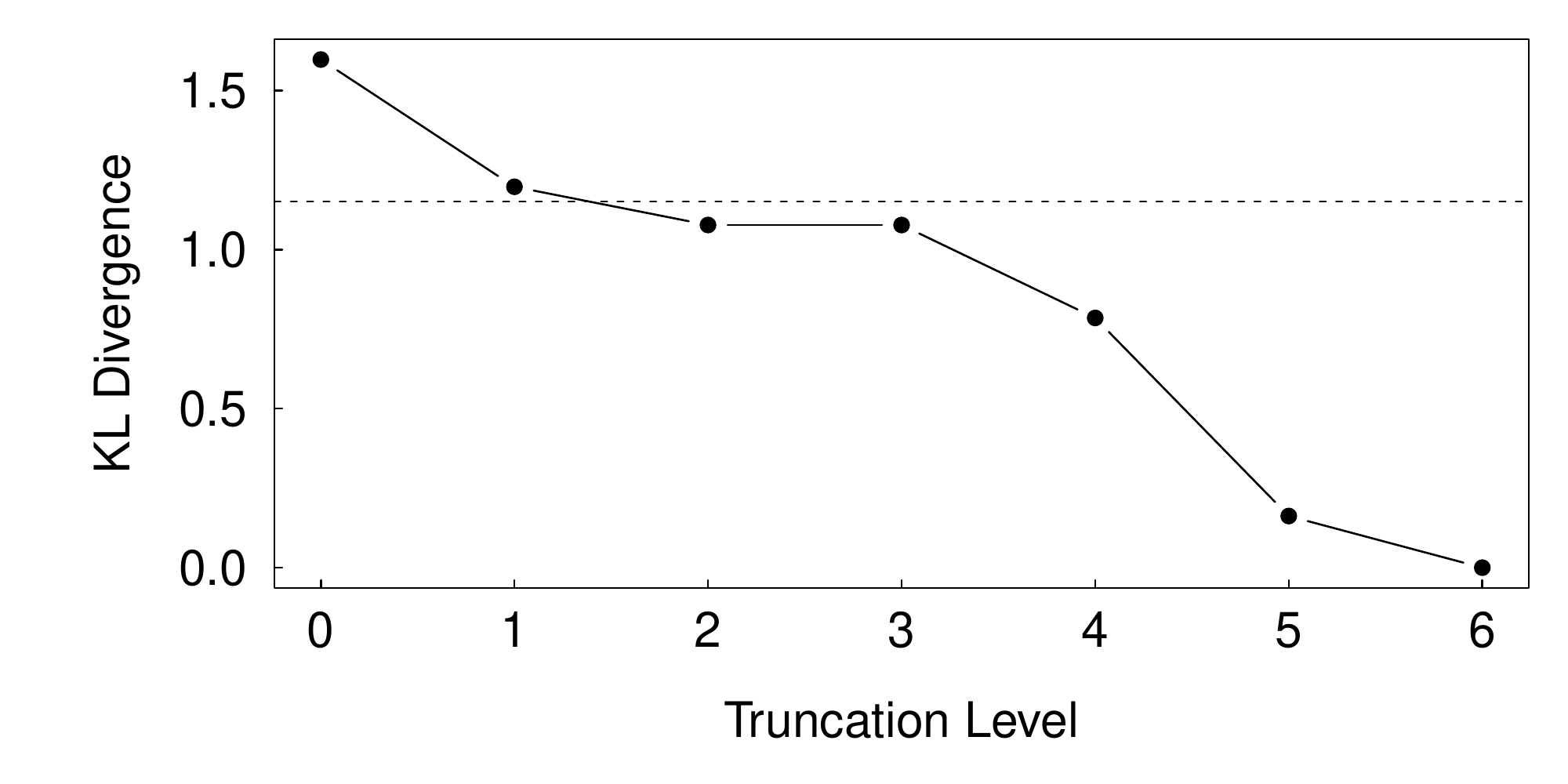}
	\caption{Kullback-Leibler divergence between the full dependence model and all possible truncated canonical vine dependence models. Values above the horizontal dashed line indicate substantial evidence \protect \citep[sensu][]{Jeffreys1961} against truncation.}\label{fig:KL}
\end{figure}

The  elicitation of dependencies for the systematic component provided substantial information. The amount of information lost by reporting a truncated canonical vine dependence model $H'$ is assessed in Figure \ref{fig:KL} by comparison with the elicited model $H$ that has no truncation. The Kullback-Leibler divergence between $H$ and $H'$ for the joint distribution of the unknown systematic component and dispersion (see Section \ref{sec:seagrassKL} for details) is interpreted as an expected log Bayes factor in favour of $H$ and against $H'$ given that $H$ is true. The categories for evidence suggested by \citet[][Appendix B]{Jeffreys1961} indicate ``substantial''  evidence against a vine dependence model truncated after level $\mathcal{T}_0$ (Figure \ref{fig:KL}) that omits the elicited canonical vine altogether. Truncation after levels $\mathcal{T}_{1:3}$ results in values near the threshold of $\log \sqrt{10} \approx 1.15$ for substantial evidence. 

\section{Conclusion}\label{sec:discussion}

Expert assessments are made measurable and accessible to scientific investigation in the Bayesian approach. To support this aspiration,  the statistical elicitation method proposed here provides a unified and flexible framework for informative prior specification of GLMs and extensions that apply to a wide array of data types and applications. The elicitation method is general and accommodates various  strategies designed to facilitate probabilistic expert elicitation. 

For example, the development of graphical feedback approaches may usefully enhance an expert's understanding of the information contained within a sample mean via the central limit theorem \citep[e.g.,][]{Dinov2008, Zhang2022}, and so assist the facilitated elicitation of an unknown dispersion parameter in Section \ref{sec:dispersion_EDM}. Advances in the communication of statistical concepts may also assist elaborations here. For instance, different choices of exponential dispersion models by Eq. \eqref{eq:sample_mean} induce alternative  approximating t distributions for the sample mean through the corresponding variance functions. Sample mean predictions elicited across multiple realisations of the systematic component could then be evaluated for consistency with each of the models, and so suggest preferences for $f$ in Eq. \eqref{eq:model_obs}. The general elicitation method also allows for the choice of feedback techniques that present quantiles to the expert. The optimal choice often depends on the domain, time availability and background of the expert \citep[e.g.,][]{Kadane1998, Garthwaite2005, OHagan2006book}.  The statistical elicitation method for GLMs and extensions therefore allows for the choice of quantiles to present as feedback to the expert. 

A consistent and recurring theme among recommendations for statistical elicitation methods is the need for software to support application \citep[][]{Kadane1980, Garthwaite2013, Elfadaly2020, Mikkola2021}. The procedures that support the statistical elicitation method for Bayesian GLMs and extensions described here are implemented in the \textsf{R} package \textbf{eglm} \citep{Hosack2023}.

\section*{Acknowledgments}
	The author thanks Keith Hayes, David Peel, two anonymous reviewers and associate editor for their constructive comments on a previous version of the manuscript. The author is grateful for the contribution of scientific expertise provided by Professor Gregory Jenkins, School of BioSciences, University of Melbourne, in the elicitation case study that was conducted with ethics approval from the CSIRO Social Science Human Research Ethics Committee (application number 103/20).

\phantomsection

\appendix
\appendixpage

\SetAlgorithmName{Procedure}{procedure}{List of Procedures}

\setcounter{equation}{0}
\setcounter{table}{0}
\setcounter{section}{0}
\setcounter{algocf}{0}
\setcounter{figure}{0}

\renewcommand*{\thesection}{S\arabic{section}}
\renewcommand*{\theequation}{S\arabic{equation}}
\renewcommand*{\thetable}{S\arabic{table}}
\renewcommand*{\thefigure}{S\arabic{figure}}

\renewcommand*\thealgocf{S\arabic{algocf}}

\section{t Distributions}\label{sec:t}

The generalised multivariate t distribution \citep[][Ch. 5]{Kotz2004} has the representation $z \sim St_p(\delta, \Sigma, r, s)$ if for $p \times 1$ location vector $\delta$ and $p\times p$ covariance matrix $\Sigma$,
\begin{equation*}
	z = \delta + v^{-1/2} y\,,\quad y \sim N(0, \Sigma)\,, \quad v \sim G(s/2, r/2)\,,
\end{equation*}
where $\mathbb{E}[v] = s/r$  so that $rv/s \sim G(s/2, s/2)$. An equivalent parametrisation is
\begin{equation*}
	z = \delta + v^{-1/2} y\,,\quad y\sim N(0, (r/s)\Sigma)\,, \quad v\sim G(s/2, s/2)\,.
\end{equation*}
The latter parametrisation is equivalent to the multivariate t distribution, see \citet[][Ch. 1]{Kotz2004} or \citet[][Ch. 2]{Joe2015}, with density
\begin{equation}
	t_p(z \mid \delta, (r/s)\Sigma, s) =  \frac{\Gamma((s + p)/2)}{\Gamma(s/2)(\pi s)^{p/2}}\left|(r/s)\Sigma\right|^{-\frac{1}{2}}\left[1 + \frac{1}{s}(z - \delta)^\top (s/r)\Sigma^{-1}\left(z - \delta\right)\right]^{-\frac{s + p}{2}}.\label{eq:t}
\end{equation}
By  Eqs. \eqref{eq:beta_marginal} and \eqref{eq:t}, the density is equivalently expressed by $z \sim t_p(\delta, (r/s)\Sigma, s) = St_p(\delta, (r/s)\Sigma, s, s) = St_p(\delta, \Sigma, r, s)$. The median of $z_i$ is $\delta_i$, which is also the mean of $z_i$ if $s > 1$. As the parameter $s \rightarrow \infty$ then the multivariate t approaches the multivariate normal with location $m$ and covariance matrix $\Sigma$ \citep[][Ch. 1]{Kotz2004}. The limiting multivariate normal for the generalised t is obtained by setting $r = ks$ for $k >0$ then letting $s\rightarrow\infty$ so that $z\sim St_p(\delta, \Sigma, ks, s) = St_p(\delta, k\Sigma, s, s) = t_p(\delta, k\Sigma, s) \rightarrow N(\delta, k\Sigma)$. The univariate t distribution with zero location and scale of one is equivalent to the Student's t distribution.

\section{Elicitation of Prior for Unknown Dispersion}

\subsection{Berry-Esseen Bound and Log Convexity Inequality}\label{sec:BEbound}

Guidance for the choice of sample size $w$ may be obtained by consideration of the Berry-Esseen bound. If the third absolute central moment of $\psi \mid \hat{\mu}_0, \lambda$ is finite, then a bound on the approximation error of Eq. \eqref{eq:CLT} is available by Eq. \eqref{eq:BE}. The first inequality in Eq. \eqref{eq:BE} uses the Berry-Esseen inequality for i.i.d. random variables. This inequality conditions on $\hat{\mu}_0$, $\phi$ and $v(\cdot)$ in Eq. \eqref{eq:BE}. For application, the facilitator would be required to hypothesise how the expert's third absolute central moment $\breve{\mu}^3$, given the choice of $\hat{\mu}_0$, might behave if the dispersion $\phi$ is known or fixed to a best estimate. This task may pose a challenge for intuition and introduce the need for additional calculations \citep[e.g.,][]{Zhang2022}. 

Alternatively, the second inequality of Eq. \eqref{eq:BE} provides an upper bound in terms of the kurtosis, $\breve{\mu}^4$. Analytical expressions of this measure of concentration in a distribution's tails, or instead the related excess kurtosis defined by $\breve{\mu}^4 - 3$, are often presented in summaries of distributional properties for exponential dispersion models with fixed $\phi$. For a nonnegative random variable $z$, the log convexity inequality of Lyapunov \citep[see][]{DasGupta2008} is
\begin{equation*}
	\mathbb{E}\left[z^{\alpha_2}\right]\mathbb{E}\left[z^{\beta - \alpha_2}\right] \leq
	\mathbb{E}\left[z^{\alpha_1}\right]\mathbb{E}\left[z^{\beta - \alpha_1}\right]\,, \quad 0 \leq \alpha_1 < \alpha_2 \leq \beta/2\,.
\end{equation*}
For a real-valued random variable $\psi$ with finite variance $\sigma^2 = \mathbb{V}[\psi] < \infty$, let $\mu = \mathbb{E}[\psi]$, $z = (|\psi - {\mu}|)/\sigma$, $\alpha_1 = 2$, $\alpha_2 = 3$ and $\beta = 6$ in the above inequality to obtain
\begin{equation*}
	\left(\breve{\mu}^3\right)^2 = \left(\frac{\mathbb{E}\left[|\psi - \mu|^3\right]}{\sigma^3}\right)^2 \leq \frac{\mathbb{E}\left[|\psi - \mu|^4\right]}{\sigma^4} = {\breve{\mu}^4}\,.
\end{equation*}
The second bound of Eq. \eqref{eq:BE} follows for $\mu = \hat{\mu}_0$ and $\psi \mid \hat{\mu}_0, \phi \sim ED(\hat{\mu}_0, \phi)$ with  variance $\mathbb{V}[\psi \mid \hat{\mu}_0, \phi] = \phi v(\hat{\mu}_0)$.

\subsection{Approximate Distribution of Sample Mean}\label{sec:discrepancy}

Given the ability to draw Monte Carlo samples from the exponential dispersion model $f = ED(\mu, \phi)$, which is a natural exponential family model conditional on $\phi$, the following procedure is used to generate $n$ samples from the joint distribution of the sample mean and an unknown index parameter $\lambda$. Given the parameters $s$ and $r$ in Eq. \eqref{eq:model_lambda} and the choice of $\hat{\mu}_0$ and $w$, for $i = 1, \ldots, N$:
\begin{enumerate}
	\item Draw realisation $\lambda_i \mid s, r \sim G(s/2, r/2)$ from Eq. \eqref{eq:model_lambda}.
	\item Draw realisation $\bar{\psi}_i \mid \hat{\mu}_0, \lambda_i, w \sim ED(\hat{\mu}_0, 1/(w\lambda_i))$ by Eq. \eqref{eq:EDmean}.
\end{enumerate}
The above steps generate $N$ paired samples $(\bar{\psi}_i, \lambda_i )$ from the joint distribution of $\bar{\psi}$ and $\lambda$ described by $f$ and $G$.

A Monte Carlo estimate of the cumulative distribution function of the sample mean,  $\hat{F}(\bar{\psi} \mid \hat{\mu}_0, w, s, r)$, is constructed based on the empirical cdf of the $N$ samples for the sample mean $\bar{\psi}$. This estimate for the distribution ${F}(\bar{\psi} \mid \hat{\mu}_0, w, s, r)$ includes Monte Carlo error. The Dvoretzky, Kiefer and Wolfowitz inequality \citep[see][Theorem 11.6]{Kosorok2008} gives
\begin{equation*}
	P\left(\sup_{x\in \mathbb{R}} \sqrt{N}\left|\hat{F}(\bar{\psi} \mid \hat{\mu}_0, w, s, r) - {F}(\bar{\psi} \mid \hat{\mu}_0, w, s, r)\right| > \epsilon \right) \leq 2\exp[-2\epsilon^2]\,.
\end{equation*}
A simultaneous confidence interval of probability $1 - \alpha$ is then
\begin{equation*}
	\hat{F}(\bar{\psi} \mid \hat{\mu}_0, w, s, r) - \epsilon \leq {F}(\bar{\psi} \mid \hat{\mu}_0, w, s, r) \leq \hat{F}(\bar{\psi} \mid \hat{\mu}_0, w, s, r) + \epsilon\,, \quad \epsilon = \sqrt{-\log(\alpha/2)/(2N)}\,.
\end{equation*}
The estimated cdf $\hat{F}(\bar{\psi} \mid \hat{\mu}_0, w, s, r)$ and its double-sided confidence interval is then graphically compared with the approximating distribution function produced by the elicitation, $T_1([\bar{\psi} - \hat{\mu}_0]/\sqrt{v_{\phi}(\hat{\mu}_0, w)} \mid s)$, where $T_1(\cdot)$ is the distribution function of the Student's t. The approximating t distribution corresponds to replacing the  exponential dispersion model $f(\mu, \phi/w) = ED(\mu, \phi/w)$  with the  normal distribution $N(\mu, \phi/w)$. Both choices describe the index parameter $\lambda = 1/\phi$ via the gamma distribution $G(\lambda \mid s/2, r/2)$ by  Eq. \eqref{eq:model_lambda}.  A numerical measure of discrepancy is provided by the Kolmogorov distance of Eq. \eqref{eq:KD}, which may be compared against the numerical error bound $\epsilon$. The latter is reduced by increasing the Monte Carlo sample size $N$.

\subsection{Kullback-Leibler Divergence for Sample Mean}\label{sec:KLdisp}

Let $F_{\bar{\psi}}$ and $G_{\lambda}$ denote the distribution functions of the conditional sample mean ($\bar{\psi} \mid \hat{\mu}_0, \lambda$) and the index parameter ($\lambda$), respectively. The Monte Carlo samples from Section \ref{sec:discrepancy} allow an estimate of the Kullback-Leibler divergence from Eq. \eqref{eq:KLdisp},
\begin{align*}
	D_{\bar{\psi}, \lambda} &= 
	\int\int \log \frac{f(\bar{\psi} \mid \hat{\mu}_0, 1/(w \lambda))G(\lambda \mid s/2, r/2)}{N(\bar{\psi} \mid \hat{\mu}_0, 1/(w \lambda))G(\lambda \mid s/2, r/2)}dF_{\bar{\psi}}(\bar{\psi} \mid \hat{\mu}_0, 1/(w \lambda))dG_{\lambda}(\lambda \mid s/2, r/2)\\
	&=
	\int\int \log \frac{f(\bar{\psi} \mid \hat{\mu}_0, 1/(w \lambda))}{N(\bar{\psi} \mid \hat{\mu}_0, 1/(w \lambda))}dF_{\bar{\psi}}(\bar{\psi} \mid \hat{\mu}_0, 1/(w \lambda))dG_{\lambda}(\lambda \mid s/2, r/2)\\
	&\approx \frac{1}{N} \sum_{i = 1}^N \log \frac{f(\bar{\psi}_i \mid \hat{\mu}_0, 1/(w \lambda_i))}{N(\bar{\psi}_i \mid \hat{\mu}_0, 1/(w \lambda_i))} = \hat{D}_{\bar{\psi}, \lambda}^N\,.
\end{align*}
Confidence bounds for the Monte Carlo error \citep[][Ch. 3]{Robert2005} are available based on the variance estimate
\begin{equation*}
	\hat{v}_N = \frac{1}{N^2}\sum_{i = 1}^N \left[\log \frac{f(\bar{\psi}_i \mid \hat{\mu}_0, 1/(w \lambda_i))}{N(\bar{\psi}_i \mid \hat{\mu}_0, 1/(w \lambda_i))} - \hat{D}_{\bar{\psi}, \lambda}^N\right]^2
\end{equation*}
by the approximation $({\hat{D}_{\bar{\psi}, \lambda}^N - D_{\bar{\psi}, \lambda}})/{\hat{v}_N^{1/2}} \sim N(0, 1)$ for Monte Carlo sample size $N\rightarrow \infty$. The  divergence $D_{\bar{\psi}, \lambda}$ is the discrepancy from the true distribution to the approximating normal-gamma distribution  for the joint distribution of the sample mean and index parameter $\lambda$. Given equal prior odds for the choice of the true model versus its approximation, the above directed divergence is the expected logarithm of the posterior odds in favour of the true model and against the approximation.

\section{Systematic Component}

\subsection{Univariate Density and Distribution Functions}\label{sec:cond_dists}

Eq. \eqref{eq:mu} gives the marginal distribution of the systematic component (Section \ref{sec:marginal}). The marginal density of the systematic component for scenario $u_i^\top$ is 
\begin{equation}
	f\left(\mu_i \mid m_{i}, V_{i, i}, r, s\right) = St_1\left(g(\mu_i) \mid m_{i}, V_{i, i}, r, s\right)\,|dg(\mu_{i})/d\mu_{i}|\,.\label{eq:mu_density}
\end{equation}
The univariate marginal distribution function of the systematic component is
\begin{equation}
	F(z \mid m_i, V_{i,i}, r, s) = T_1\left(
	\left.	\frac{g(z) - m_{i}}{(r/s)^{1/2}V_{i,i}^{1/2}}\,
	\textrm{sgn}\frac{dg}{d\mu} \right. 
	\mid s
	\right)\, , \quad z \in \Omega\, ,\label{eq:mu_dist}
\end{equation}
where $T_1(z \mid s)$ is the cumulative distribution function of the Student's t and $\textrm{sgn}\, h$ is the signum function that equals $+1$ if $h > 0$ and $-1$ if $h < 0$. The sign of $\textrm{sgn}\,{dg}/{d\mu}$ is constant within the mean domain because the link function $g$ is continuous and invertible (Section \ref{sec:glm}). The corresponding quantile function is
\begin{equation}
	F^{-1}(q \mid m_i, V_{i,i}, r, s) = g^{-1}\left(m_i + (r/s)^{1/2}V_{i,i}^{1/2}T_1^{-1}(q \mid s)\,
	\textrm{sgn}\,\frac{dg}{d\mu}\right)\,,\quad q\in(0, 1)\,.
	\label{eq:mu_q}
\end{equation}

Eq. \eqref{eq:cond_eta} in Section \ref{sec:DCMP} provides the conditional density of the linear predictor given realised values of the linear predictor. The univariate conditional distributions are available and may be used to provide graphical and numerical feedback to the expert at tree level $\mathcal{T}_l$ (Section \ref{sec:estimation}). For link function $g$, the univariate conditional density of the systematic component at scenario $u_k^\top$ given realised values of the systematic component $\hat{\mu}_{1:l}$ at the first $l$ scenarios is
\begin{equation}
	f\left(\mu_k \mid m_{k \mid 1:l}, V_{k, k \mid 1:l}, r, s\right)  = St_1(g(\mu_k) \mid m_{k \mid 1:l},  V_{k, k \mid 1:l, 1:l}, r + \zeta(\hat{\eta}_{1:l}), s + l) \,|dg(\mu_{k})/d\mu_{k}|\,,\label{eq:mucond_density}
\end{equation}
where $k > l$, $\hat{\eta}_{1:l} = g(\hat{\mu}_{1:l})$ and $\zeta(\hat{\eta}_{1:l}) = (\hat{\eta}_{1:l} - m_{1:l})^\top V_{1:l, 1:l}^{-1}(\hat{\eta}_{1:l} - m_{1:l})$. The univariate conditional distribution function of the systematic component at scenario $u_k^\top$ conditional on realised values of the systematic component $\hat{\mu}_{1:l}$ at the first $l$ scenarios is
\begin{align}
	F_{\mu}\left(z \mid m_{k \mid 1:l}, V_{k, k \mid 1:l}, r, s\right)  
	&= T_1\left(
	\left.	\frac{g(z) - m_{k \mid 1:l}}{\left[(r + \zeta(\hat{\eta}_{1:l}))/(s + l)\right]^{1/2}V_{k,k \mid 1:l}^{1/2}} \,
	\textrm{sgn}\frac{dg}{d\mu} \right. \mid s + l
	\right)\,, \label{eq:mucond_dist}
\end{align}
where $z \in \Omega$. The conditional quantile function with $q\in(0, 1)$ is
\begin{align}
	F_{\mu}^{-1}\left(q \mid m_{k \mid 1:l}, V_{k, k \mid 1:l}, r, s\right)
	&= g^{-1}\left[m_{k|1:l} + \sqrt{\frac{[r + \zeta(\hat{\eta}_{1:l})]V_{k,k \mid 1:l}}{s + l}}T_1^{-1}\left(q \mid s + l\right)\,
	\textrm{sgn}\,\frac{dg}{d\mu}\right]\,.\label{eq:mucond_q}
\end{align}

\subsection{Algorithm S1 for $P \rightarrow R$ Mapping and Bounds for Conditional Medians}\label{sec:bounds_median}

\SetAlgoProcName{Algorithm}{Algorithm}

\begin{procedure}[ht!]
	\SetAlgoLined
	\KwData{$n\times n$ partial correlation matrix $\Rho = \left[p_{lk}\right]$ where $p_{lk} = \rho_{l,k|1:(l-1)}$ for $l < k$}
	\KwResult{$n\times n$ correlation matrix $R = \left[\rho_{l,k}\right]$}
	initialise $\rho_{1k} = \rho_{k1} \leftarrow p_{1k}$ for $k = 2:n$ \;
	\If{$n > 2$}{
		\For{$l = 2:(n-1)$}{
			\For{$k = (l+1):n$}{
				$r \leftarrow p_{lk}$ \;
				\For{$\gamma = (l-1):1$}{
					$r\leftarrow p_{\gamma l}p_{\gamma k} + r(1 - p_{\gamma l}^2)^{1/2}(1 - p_{\gamma k}^2)^{1/2}$ \;
				}
				$\rho_{l,k} = \rho_{k,l} \leftarrow r$ \;
			}
		}
	}
	\Return matrix $R = \left[\rho_{l,k}\right]$ with entries of main diagonal equal to one.
	\caption{S1(). Mapping from partial correlation matrix $\Rho$ to correlation matrix $R$ for canonical vine, see Algorithm 27 of \citet{Joe2015}.}\label{alg:PR}
\end{procedure}

\citet{Joe2015} provides the mapping $\Rho \rightarrow R$ shown in Algorithm \ref{alg:PR}.
At the start of the elicitation for scenario $u_{k}^\top$ at tree level $\mathcal{T}_{l}$ with $ 1 \leq l < k$, let the partial correlation matrix $\Rho$ with the zero entry $\Rho_{l,k}$ replaced by $\rho \in \{-1, + 1\}$ be denoted by $\Rho^{(\rho)}$. Let $R^{(\rho)}$ denote the corresponding correlation matrix induced by Algorithm \ref{alg:PR} applied to $\Rho^{(\rho)}$. The corresponding scale matrix for the linear predictor is $V^{(\rho)} = \diag[V]^{1/2}R^{(\rho)}\diag[V]^{1/2}$, where the diagonal entries of the scale matrix were previously elicited in Section \ref{sec:marginal} such that $\diag[V^{(\rho)}] = \diag[V]$. From Eq. \eqref{eq:median_diff}, the feasible bounds for $c_{k|1:l}$ are given by the range of
\begin{equation*}
	g^{-1}\left(m_k + V^{(\rho)}_{k, 1:l} \left[V^{(\rho)}_{1:l, 1:l}\right]^{-1}\left(\hat{\eta}_{1:l} - m_{1:l}\right)\right)\textrm{ for } \rho \in \{\pm 1\}\,.
\end{equation*}
These bounds on the conditional median are available for graphical or numerical presentation to the expert.

\section{Induced Prior}\label{sec:KL}

Section \ref{sec:induced-prior} presents the elicited parameters $\theta = \left\{s, r, m, V\right\}$ for the saturated model $H$  described by Eqs. \eqref{eq:model}  and \eqref{eq:eta}. The choice of observation model $f' \in \mathcal{F}$ in Eq. \eqref{eq:model_obs} and model matrix $X \in \mathcal{X}$ of full column rank in Eq. \eqref{eq:model_beta} induces a new alternative model $H'$ with parameters $\theta' = \{s', r', \delta', \Sigma'\}$.
The conditional property of information \citep[][Ch. 2]{Kullback1959} applied to Eq. \eqref{eq:KLparam} obtains
\begin{equation}
	D(H:H')
	= \int D\left(H:H' \mid \lambda\right) G\left(\lambda \mid \frac{s}{2}, \frac{s}{2}\right)d\lambda + \int \log \frac{G\left(\lambda \mid \frac{s}{2}, \frac{s}{2}\right)}{G\left(\lambda \mid \frac{s'}{2}, \frac{s'}{2}\right)} G\left(\lambda \mid \frac{s}{2}, \frac{s}{2}\right)d\lambda\,\label{eq:conditional_KL}
\end{equation}
where 
\begin{align}
	D\left(H:H' \mid \lambda\right)
	&=
	\int \log \frac{N\left(\eta \mid Am, \frac{r}{s\lambda}AVA^\top\right)}{N\left(\eta \mid X\delta', \frac{r'}{s'\lambda}X\Sigma'X^\top\right)} N\left(\eta \mid Am, \frac{r}{s\lambda}AVA^\top\right) d\eta\nonumber\\
	&= \frac{1}{2}\log \frac{\left|S'\right|}{\left|S\right|} + \frac{1}{2}\mathop{\mathrm{Tr}}\left[ S\left(S'^{-1} - S\right)\right] + \frac{\lambda}{2}\mathop{\mathrm{Tr}} \left[S'^{-1}(\gamma - \gamma')(\gamma - \gamma')^\top\right]\,,\label{eq:KL_normal}
\end{align}
is the Kullback-Leibler divergence of a multivariate normal \citep[][Ch. 9]{Kullback1959} with
\begin{equation*}
	\gamma = Am\,, \quad \gamma' = X\delta'\,, \quad S = (r/s)AVA^\top\,, \quad S' = (r'/s')X\Sigma'X^\top\,.
\end{equation*}
With respect to $\gamma$ and $\gamma'$, Eq. \eqref{eq:KL_normal} is minimised for the substitutions $A = \hat{A}$ and $\delta' = \hat{\delta}$ from Eq. \eqref{eq:parms}, whereby $\gamma = \gamma'$. Then the last addend in Eq. \eqref{eq:KL_normal} is zero and $D(H:H' \mid \lambda)$ does not depend on $\lambda$. Thus, with respect to $s'$, $D(H:H')$ is minimised by the substitution of $s' = {s}$ so that the second integral of Eq. \eqref{eq:conditional_KL} is zero. Next, note that the substitutions  $\Sigma' = \hat{\Sigma}$, $s' = \hat{s}$ and $r' = \hat{r}$ into $S'$ and $A = \hat{A}$ into $S$ obtain the result $S = S' = (r/s)X\left(X^\top V^{-1} X\right)^{-1}X^\top$. Eqs. \eqref{eq:conditional_KL} and \eqref{eq:KL_normal} are therefore zero given the substitutions from Eq. \eqref{eq:parms}. The result $D(H:H') = 0$ is the minimum by the convexity property of the Kullback-Leibler divergence \citep[][Ch. 2]{Kullback1959}. 

If $p < n$ and, contrary to Eq. \eqref{eq:parms}, the identity map $A = I$ is assumed then $D(H:H') = \infty$ because $X\beta$ in model $H'$ lies in a subspace of the saturated model $H$. The model $H'$ would then contradict the expert assessments documented by $H$. If instead the assumption that $A$ is a projection matrix onto $\mathbf{C}(X)$ is accepted, then the optimal projection is $\hat{A}$ and the divergence is minimised such that $D\left(H : H'\right) = 0$. 

If $p = n$ then the optimal projection is the identity  $\hat{A} = I$  with $\hat{\delta} = X^{-1}m$ and $\hat{\Sigma} = X^{-1} V X^{-\top}$ by Eq. \eqref{eq:parms}. Specifying a square model matrix $X$ of full rank ensures a non-singular transformation between $\beta$ and the elicited systematic component $\mu$ with no loss of information. 

For the case of conditional means priors (CMPs) where $p < n$, \citet[][]{Bedrick1996} discuss how an expert is unlikely to provide responses within $\textbf{C}(X)$. For a CMP with known dispersion in the normal linear model, \citet[][]{Bedrick1996} derive $\hat{\delta}$ and $\hat{\Sigma}$ in Eq. \eqref{eq:parms} by first imagining $n - p$ predictors with associated coefficients $\beta^{\ast}$, then taking the conditional distribution of $\beta \mid \beta^\ast = 0$ as the prior. The above interpretation in terms of $D\left(H : H'\right) $ does not depend on imagined coefficients $\beta^{\ast}$ and applies to the more general Eq. \eqref{eq:model}. 

\section{Extensions}

\subsection{Variance Approximation for Standard Simplex Distribution}\label{sec:svar}

The standard simplex distribution with density
\begin{equation*}
	S(y \mid \mu, \lambda) = \left(\frac{\lambda}{2\pi\left[y(1 - y)\right]^3}\right)^{\frac{1}{2}}\exp\left[-\frac{\lambda}{2}d(y, \mu)\right]\,,  \quad y, \mu \in (0, 1)\,,
\end{equation*}
where $d(y, \mu) = \frac{(y - \mu)^2}{y(1 - y)\mu^2(1 - \mu)^2}$, is an example of a dispersion model available for use by an extended GLM (Section \ref{sec:DM}). A related distribution \citep[][Ch. 5]{Jorgensen1997a, Jorgensen1997} is 
\begin{equation}
	S^+(y \mid \mu, \lambda) = \frac{1}{\Gamma(1/2, \lambda/2)\sqrt{\pi e^\lambda}}\left[y(1 - y)\right]^{-\frac{1}{2}}
	\exp\left[-\frac{\lambda}{2}d(y, \mu)\right]\,,\quad y,\mu\in(0, 1)\,,\label{eq:simplex_pos}
\end{equation}
where $d(y, \mu) = (y - \mu)^2/(y(1 - y))$ and $\Gamma(a, b) = \int_b^\infty t^{a-1}e^{-t}dt$ is the (upper) incomplete gamma function. \citet[][Ch. 5]{Jorgensen1997a, Jorgensen1997} suggest calculation of the variance for $y \sim S(\mu, \lambda)$ by use of the mixed moments $\mathbb{E}[y^{b_1}(1-y)^{b_2}]$ (see Eq. 7.2 of \citet{Jorgensen1997a} or Eq. 5.57 of \citet{Jorgensen1997}), which can be expressed as a ratio of normalising constants of the standard simplex distribution and Eq. \eqref{eq:simplex_pos} (see Table 1 of \citet{Jorgensen1997a} or Table 5.1 of \citet{Jorgensen1997}). Applying this derivation, \citet{Jorgensen1997a} notes that the variance of the standard simplex is approximately $\mathbb{V}[y] = \mu^3(1 - \mu)^3/\lambda$ for large $\lambda$ or small $\mu$ or $1 - \mu$. 

This approximation to the variance of the standard simplex distribution can be shown as follows. Since $\mathbb{E}[y] = \mu$ for the standard simplex distribution, applying the approach described above for the variance of the standard simplex distribution using mixed moments with $b_1 = b_2 = 1$ obtains the expression \citep[cf.][Ch. 5]{Jorgensen1997a, Jorgensen1997}, 
\begin{equation}
	\mathbb{V}[y] = \mu(1 - \mu) - \left(\frac{\lambda}{2}\right)^{1/2}\exp\left[\frac{\lambda}{2\mu^2(1 - \mu)^2}\right]\Gamma\left(\frac{1}{2}, \frac{\lambda}{2\mu^2(1 - \mu)^2}\right)\,.\label{eq:variance}
\end{equation}
The equivalence $\Gamma(1/2, z) = \sqrt{\pi} \textrm{erfc}(\sqrt{z})$ \citep[][Eq. 8.4.6]{NIST2020}, where $\textrm{erfc}(z)$ is the complementary error function, suggests for large $\lambda$ or small $\mu$ or $1 - \mu$ the approximation 
\begin{align*}
	\Gamma\left(\frac{1}{2}, \frac{\lambda}{2\mu^2(1 - \mu)^2}\right) &= \sqrt{\pi} \textrm{erfc}\left(\sqrt{\frac{\lambda}{2\mu^2(1 - \mu)^2}}\right) \\
	&\approx \left(\frac{2}{\lambda}\right)^{1/2}\exp\left[-\frac{\lambda}{2\mu^2(1 - \mu)^2}\right]\left\{\mu(1 - \mu) - \left(\frac{\mu^3(1 - \mu)^3}{\lambda}\right) \right\}\,.
\end{align*}
The above approximation uses the first two terms of an asymptotic expansion of the complementary error function $\textrm{erfc}(z)$  that holds as $z \rightarrow \infty$ \citep[][Eq. 7.12.1]{NIST2020}. Substituting the above approximation into Eq. \eqref{eq:variance} yields the approximation $\mathbb{V}[y] \approx v(\mu)/\lambda$, where $v(\mu) = \mu^3(1 - \mu)^3$ is the variance function of the standard simplex distribution. This approximation holds for large  $\lambda$ or small $\mu$ or $1 - \mu$.

\subsection{Lognormal with Unknown Scale}\label{sec:lognormal}

In Section \ref{sec:over_Poisson}, the  mean of the Poisson distributed random variable $y$ given $\psi$ is $z = e^\psi$. Since  $\psi \sim N(\mu, \phi)$, the random variable $z$ follows a lognormal distribution, which is not an exponential dispersion model. But $k\log z\sim N(k\mu, k^2\phi)$, given $\mu$ and $\phi$ in Eq. \eqref{eq:model}, where the known constant $k = 1/\log B$ accounts for the desired base $B$ of the logarithm. A lognormal model is appropriate if the logarithm of the elicitation target with base $B$ is interpretable, denoted by $\psi_B = \log_B z = k\log z$. If the elicitation target $\psi_B$  is described by Eq. \eqref{eq:model} for $f$ normal and $g$ the identity link, then the marginal distribution is $\psi_B \sim St_n(X\delta_B, X\Sigma_B X^\top, r, s)$. Procedure \ref{protocol} applied to the transformed target $\psi_B$ then elicits $\delta_B$, $\Sigma_B$, $s$ and $r$. The induced location and conditional scale for $\psi$ in Eq. \eqref{eq:model_beta} are then $\delta = \delta_B/k$ and $\Sigma = \Sigma_B/k^2$, and  $\lambda \sim G(s/2, r/2)$ in Eq. \eqref{eq:model_lambda}.

\section{Elicitation Case Study}\label{sec:seagrass}

In the elicitation case study of Section \ref{sec:study}, the potential response of annual mean \textit{Zostera nigricaulis} percent cover to changes in nitrogen levels and suspended sediments is examined for a seagrass ecosystem within Port Phillip Bay, Australia. 
The linear predictor 
\begin{align}
	\eta_i =& \beta_0 + \beta_1 \log_{10} DIN_i + \beta_2 TSS_i + \beta_3 \log_{10} DIN_i \times TSS_i + \beta_4 (\log_{10} DIN_i)^2  \nonumber \\
	&+ \beta_5 TSS_i^2 + \beta_6 (\log_{10} DIN_i)^2 \times TSS_i^2\label{eq:linpred}
\end{align}
allowed for interactions and quadratic responses with the base 10 logarithm of the annual average $DIN$ in the bottom 1 metre of the water column, and the annual average $TSS$ in the water column. 
A $n\times n$ model matrix $X$ of full column rank with the polynomial basis expansion for Eq. \eqref{eq:linpred} was derived from the $n = 7$ elicitation scenarios in the elicitation scenario matrix $U$, see Table \ref{tab:cond}.
The $i$\textsuperscript{th} row of $X$ is a design point $x_i^\top$ that corresponds to the $i$\textsuperscript{th} scenario in $U$. 

\begin{table}[h]
	\centering
	\caption{The elicitation scenarios of $U$ defined by the annual averages of dissolved inorganic nitrogen in the bottom 1 metre of the water column ($DIN$ in mg per litre) and total suspended solids in the entire water column ($TSS$ in mg per litre).   \\}\label{tab:cond}
	
	\begin{tabular}{c|cc}
		\hline
		Scenario & $DIN$ & $TSS$ \\ 
		\hline
		1 & 0.0001 & 0.1000 \\ 
		2 & 0.0500 & 0.1000  \\ 
		3 & 0.5000 & 0.1000  \\ 
		4 & 0.0001 & 12.2500  \\ 
		5 & 0.0001 & 50.0000  \\ 
		6 & 0.0500 & 50.0000  \\ 
		7 & 0.5000 & 50.0000 \\ 
		\hline
	\end{tabular}
\end{table}

The elicitation session occurred by video conference in November 2020. At the start of the elicitation session, the contributing  expert with  scientific expertise of seagrass ecosystems in coastal embayments of southeastern Australia was educated in subjective probability \citep[sensu][Ch. 3]{Lindley2014}, cognitive biases \citep[e.g.,][]{Tversky1974} and engaged in practice elicitation exercises \citep{OHagan2006book}. The definition for the quantity of interest of the elicitation (Section \ref{sec:study}) was reviewed along with the choice of the covariates that together defined the elicitation scenarios in $U$ (Table \ref{tab:cond}). 

\subsection{Random Component}\label{sec:seagrassRandom}

The elicitation of the unknown dispersion parameter (Section \ref{sec:unknown_dispersion}) determines the amount of overdispersion in the binomial response for the seagrass study.  Central credible intervals of probabilities $1/3$ and $0.90$ were elicited for the sample mean $\bar{\psi}$ conditioned on values of the systematic component $\hat{\mu}_0$ emphasising small values from $0.01$ to $0.10$  with possible sample sizes explored for $w$ between 10 and 800 quadrats. The elicited values of $s$ and $r$ were compared across the different choices of $\hat{\mu}_0$ and $w$ and iteratively modified. The final elicited parameters for the gamma prior of the index parameter $\lambda$ were $s = 14.3$ and $r = 118$. The final scenario evaluated the sample mean prediction for a percent cover of 1\% (corresponding to $\hat{\mu}_0 = 0.01$) with a relatively small sample size $w = 10$. The elicited random component reflected observation uncertainty introduced by localised disturbance events and the patchy distribution of the seagrass.

\begin{figure}[ht!]
	\centering
	\begin{subfigure}[b]{0.475\textwidth}
		\includegraphics[width=\linewidth]{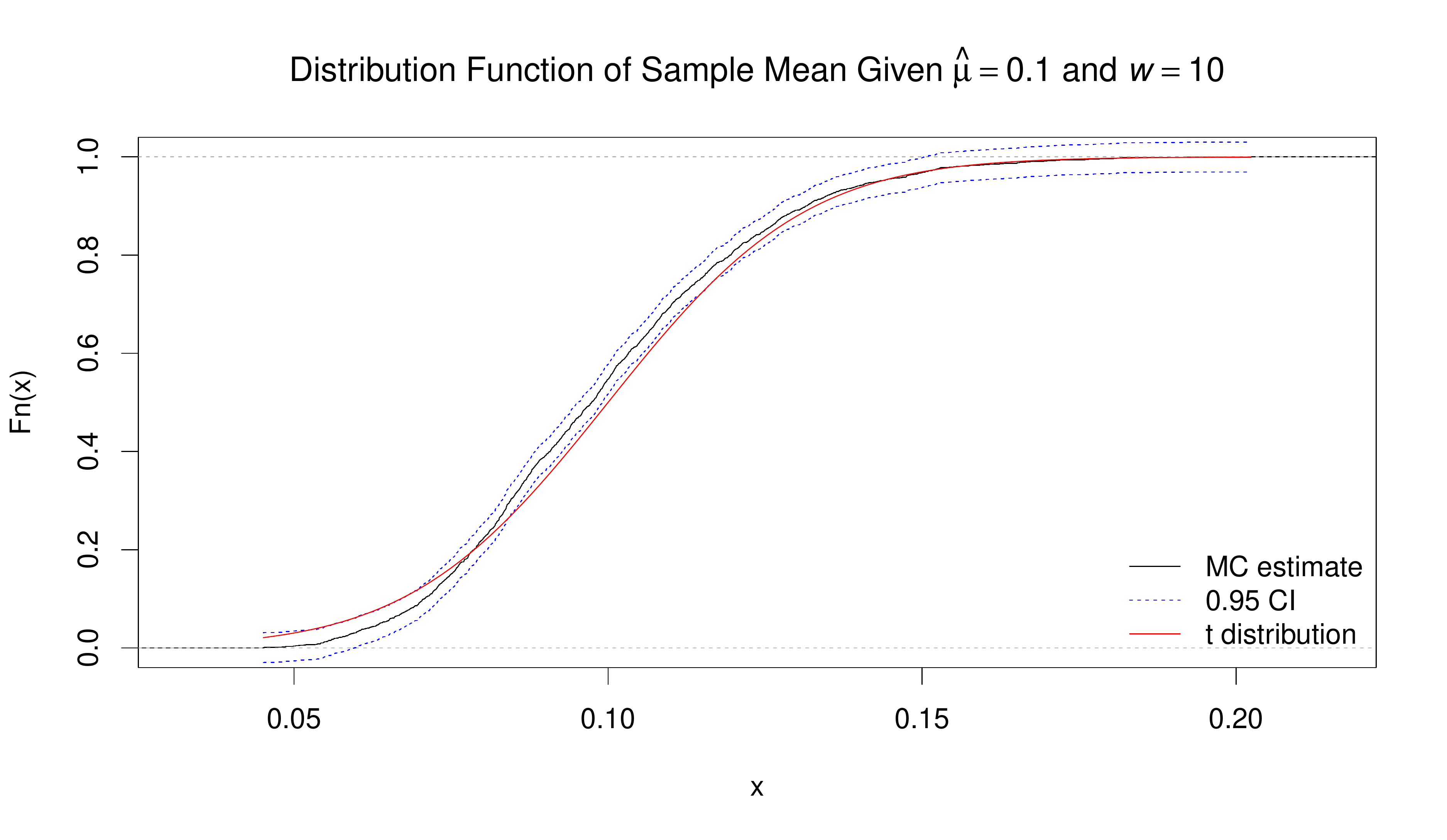}
		\caption{Penultimate scenario: $\hat{\mu}_0 = 0.10$.}
	\end{subfigure}
	\begin{subfigure}[b]{0.475\textwidth}
		\includegraphics[width=\linewidth]{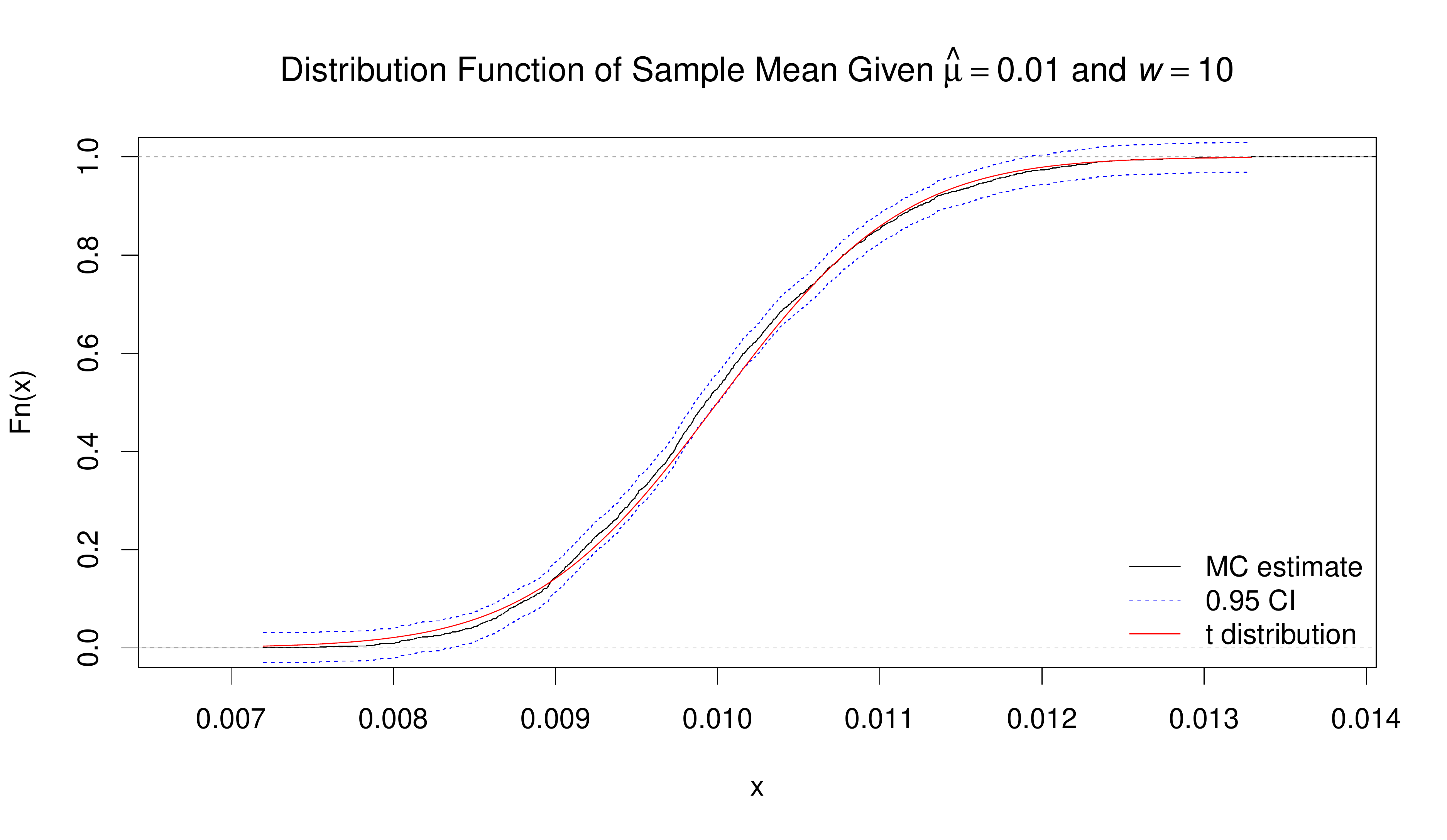}
		\caption{Final scenario: $\hat{\mu}_0 = 0.01$.}
	\end{subfigure}
	\caption{Comparisons of estimated cumulative distribution function for the sample mean against the generalised t approximation for the final two scenarios considered for the unknown dispersion with $w = 10$. The estimated distribution function and its 0.95 confidence interval are based on $N = 2000$ Monte Carlo realisations for the sample mean $\bar{\psi}$.}\label{fig:discrepancy}
\end{figure}

The discrepancies between the estimated distribution functions of the sample mean against the approximate generalised t approximation are shown for the final two choices of $w$ and $\hat{\mu}_0$. In each case, Monte Carlo realisations for $\psi$ were drawn by the inverse cumulative distribution function method for the exact Studentised simplex density \citep[][Ch. 5]{Jorgensen1997a, Jorgensen1997} derived from Eq. \eqref{eq:model_lambda} and the simplex density (Section \ref{sec:DM}),
\begin{align*}
	f(\psi \mid \mu, s, r) =& \int S(\psi \mid \mu, \lambda) G\left(\lambda \mid \frac{s}{2}, \frac{r}{2}\right)d\lambda\\
	=& \frac{\sqrt{a}}{\beta\left(\frac{b - a}{2a}, \frac{1}{2}\right)}\left[\psi(1 - \psi)\right]^{-3/2}\left[1 + a\frac{(\psi - \mu)^2}{\psi(1 - \psi)\mu^2(1 - \mu)^2}\right]^{-\frac{b}{2a}}\,,
\end{align*}
where $a = 1/r$ and $b = (s + 1)/r$. The Monte Carlo estimate and its 0.95 confidence interval were calculated according to Section \ref{sec:discrepancy} for the penultimate scenario of $\hat{\mu}_0 = 0.10$ and the final scenario of $\hat{\mu}_0 = 0.01$ for $w = 10$ (Figure \ref{fig:discrepancy}). The estimated Kolmogorov distances were $0.054$ and $0.035$, respectively, which reflects the better approximation of the generalised t for the lower value of $\hat{\mu}_0$ in the simplex distribution (Section \ref{sec:DM}).

\subsection{Systematic Component}\label{sec:seagrassSystematic}

For marginal elicitation of the systematic component (Section \ref{sec:marginal}), the numerical feedback included the scenario information encoded in the scenario matrix $U$ (Table \ref{tab:cond}), which was defined by levels of dissolved inorganic nitrogen and total suspended solids. The marginal centres and scales of the transformed generalised t distribution were elicited according to Section \ref{sec:marginal}. Graphical and numerical feedback for the elicitation target of seagrass percent cover were provided for the $1/3$ and $0.80$ central credible intervals, along with the median and plotted probability density function.

These same quantities were also presented for the elicited conditional distributions at tree levels $\mathcal{T}_{1:(n-1)}$ (Section \ref{sec:estimation}). Conditional medians were elicited where, in addition to the information in $U$, hypothetical realised values for the systematic component were introduced at tree levels greater than zero (Sections \ref{sec:DCMP} and \ref{sec:estimation}). The realised values for the systematic component $\hat{\mu}_{1:l}$ were sequentially introduced  with increasing tree level $\mathcal{T}_l$, $l = 1, \ldots, n - 1$. Thus, a new realised $\hat{\mu}_l$ was introduced for the $l$\textsuperscript{th} scenario at tree level $\mathcal{T}_l$. The conditioning values $\hat{\mu}_{1:l}$ were generated as in Section \ref{sec:conditioning}, where their tail probabilities of $(1 - \alpha)/2$ were derived from Eq. \eqref{eq:eta} given known dispersion $\lambda = 1$. This latter choice induced smaller relative changes in the conditioning values relative to their conditional medians than would have resulted from the alternative approach presented in Section \ref{sec:conditioning} because of the substantial overdispersion elicited in this case study.  Given this choice,  the conditioning values were determined by the elicited values of $m$ and $V$ in Eq. \eqref{eq:model_beta}. The magnitude of the random component was in this sense screened out while selecting conditioning values for the systematic component.

The conditioning values $\hat{\mu}_{1:l}$ were interpreted in terms of relative change with respect to previously elicited medians. During the elicitation session, the realised values of the systematic component were interpreted as proportional changes with respect to the predicted  median responses at lower tree levels. The expert judged that negligible information was provided  at tree level $\mathcal{T}_3$ for scenarios 4 to 7, and the conditional medians were unadjusted at that level. The rationale, to paraphrase, was that scenarios 1 and 3  were  predicted to have similar levels of seagrass percent cover (see Figure \ref{fig:pc}), though for different reasons: Low $DIN$ in scenario 1 suggested possible nutrient constraints on seagrass growth, whereas high $DIN$ in scenario 3 suggested enhanced growth of epiphytes and possible light suppression on seagrass. The additional realisation of scenario 3 at tree level $\mathcal{T}_3$ did not add much additional information. Scenario 4, however, had a higher amount of $TSS$ than scenarios 1 and 3 (Table \ref{tab:cond}), and the conditional medians of scenarios 5 to 7 were updated at the next tree level $\mathcal{T}_4$.

\subsection{Kullback-Leibler Divergence for Truncated Vine Models}\label{sec:seagrassKL}

The directed divergences from the full dependence model to each of the truncated dependence models are evaluated as follows. Define the truncated model $H(t)$ with scale $V(t)$ given truncated dependence after level $\mathcal{T}_t$ (Section \ref{sec:truncated-cmps}) and corresponding alternative fitted models $H'(t)$ for $t \in \{0, \ldots, n - 1\}$. From Eq. \eqref{eq:parms} and the nonsingular $X$, the divergence $D(H(t) : H'(t))$ is minimised for the choice of parameters 
\begin{equation}
	\begin{gathered}
		\hat{\delta}(t) = X^{-1} m\,, \quad  \quad \hat{\Sigma}(t) = \left(X^\top V(t)^{-1} X\right)^{-1}\,,\\
		\hat{A}(t) = I\,, \quad \hat{s} = s\,,\quad \hat{r} = r\,.
	\end{gathered}\label{eq:parmst}
\end{equation} 
Let $\hat{H}(t)$ denote the truncated model with parameters given by Eq. \eqref{eq:parmst} and model matrix $X$.
For the joint distribution of the systematic component and dispersion, the Kullback-Leibler divergence between the full dependence model $H$ described by Eqs. \eqref{eq:model} and \eqref{eq:eta} the alternative  models $\hat{H}(t)$ truncated at level $t \in \{0, \ldots, n - 1\}$ is
\begin{equation}
	\begin{aligned}
		D\left(H:\hat{H}(t)\right) =& \int D\left(H:\hat{H}(t) \mid \lambda\right)G\left(\lambda \mid \frac{s}{2}, \frac{r}{2}\right) d\lambda + \int \log \frac{G\left(\lambda \mid \frac{s}{2}, \frac{r}{2}\right)}{G\left(\lambda \mid \frac{s}{2}, \frac{r}{2}\right)} G\left(\lambda \mid \frac{s}{2}, \frac{r}{2}\right)d\lambda \\
		=& \frac{1}{2}\log \frac{\left|S'(t)\right|}{\left|V\right|} + \frac{1}{2}\mathop{\mathrm{Tr}}\left[ V\left(S'(t)^{-1} - V^{-1}\right)\right]\\ &+ \frac{s}{2r}\mathop{\mathrm{Tr}} \left[S'(t)^{-1}(m - \gamma'(t))(m - \gamma'(t))^\top\right]\,\\
		=& \frac{1}{2}\log \frac{\left|V(t)\right|}{\left|V\right|} + \frac{1}{2}\mathop{\mathrm{Tr}}\left[ V\left(V(t)^{-1} - V^{-1}\right)\right] \\
		=& \frac{1}{2}\log \frac{\left|R(t)\right|}{\left|R\right|}  + \frac{1}{2}\mathop{\mathrm{Tr}}\left[ R R(t)^{-1}\right] - \frac{n}{2}\,,
	\end{aligned}\label{eq:Dtrunc}
\end{equation}
where $\gamma'(t) = X\hat{\delta}(t) = m$ and $S'(t) = X\hat{\Sigma}(t)X^\top = V(t)$ is the scale matrix with truncated dependence and corresponding correlation matrix  $R(t)$. The above uses the conditional property of information as in Eq. \eqref{eq:conditional_KL}. Results are shown in Figure \ref{fig:KL}.

\clearpage

\section{Elicitation Procedures}\label{sec:algs}

\begin{algorithm}[ht]
	\SetAlgoLined
	\KwData{Set of exponential dispersion models $\mathcal{F}$.}
	\KwResult{$s$ and $r$ in Eq. \eqref{eq:model_lambda}; $v_{\phi}(\hat{\mu}_0, w)$ in Eq. \eqref{eq:r} if dispersion is unknown.}
	Choose $f \in \mathcal{F}$ (Section \ref{sec:components})\;
	\uIf{$\phi$ is known}{
		Let $s, r \rightarrow \infty$ such that $\lambda \triangleq 1/{\phi} = s/r$ for known dispersion ${\phi}$ (Section \ref{sec:known})\;
		\Return $s$, $r$\;
	}
	\Else{
		Choose $\hat{\mu}_0$ and sample size $w$ for sample mean $\bar{\psi} \mid \hat{\mu}_0, w$ (Section \ref{sec:dispersion_EDM})\;
		\Repeat{Approximation to sample mean is accepted}{
			\uIf{Candidate $\hat{\mu}_0$ unsatisfactory to expert}{
				Adjust $\hat{\mu}_0$\;
			}
			\uIf{Approximation to sample mean distribution unsatisfactory}{
				Increase $w$\;
			}
			Elicit lower bound $d_1$ of central credible interval with probability $\alpha_1$ for $\bar{\psi}\mid \hat{\mu}_0, w$\;
			Elicit lower bound $d_2 < d_1$ of central credible interval with probability $\alpha_2 > \alpha_1$ for $\bar{\psi}\mid \hat{\mu}_0, w$\;
			Parametrise $s$ by Eq. \eqref{eq:s} and $v_{\phi}(\hat{\mu}_0, w)$ with induced $r$ by Eq. \eqref{eq:r}\;
			Provide graphical and numerical feedback of approximate distribution via Eq. \eqref{eq:sample_mean}\;
			Assess approximation accuracy (Section \ref{sec:dispersion_EDM})\;
		}
		\Return $s$, $r$,  $v_{\phi}(\hat{\mu}_0, w)$\;
	}
	\caption{Elicitation of dispersion parameter for GLMs; see Section \ref{sec:extensions} for alternative strategies applicable to extensions of GLMs.}
	\label{alg:dispersion}
\end{algorithm}

\begin{algorithm}[ht]
	\SetAlgoLined
	\KwData{Elicitation scenario matrix $U$, link function $g$; $s$ and $r$ from Procedure \ref{alg:dispersion}.}
	\KwResult{Location vector $m$ and main diagonal of scale matrix $\diag[V]$ for the linear predictor $\eta$.}
	\For{$i = 1:n$}{
		\Repeat{expert accepts $f\left(\mu_i \mid m_{i}, V_{i, i}, r, s\right)$}{
			Elicit central credible interval $(a_i, b_i)$ of probability $\alpha$ for $\mu_i \mid u_i^\top$ (Section \ref{sec:marginal})\;  
			Parametrise  $m_i$ and $V_{ii}$ conditional on $g$, $s$, $r$, $(a_i, b_i)$ (Section \ref{sec:marginal})\;
			Provide graphical and numerical feedback via Eqs. \eqref{eq:mu_density}, \eqref{eq:mu_dist} and \eqref{eq:mu_q} \;
		}
	}	
	\Return $m$, $\diag[V]$.
	\caption{Elicitation of location vector $m$ and scale parameters $\diag[V]$ for linear predictor $\eta$.}
	\label{alg:CI}
\end{algorithm}

\begin{algorithm}[H]
	\SetAlgoLined
	\KwData{Elicitation scenario matrix $U$ and link function $g$;  $s$ and $r$ from Procedure \ref{alg:dispersion}; location vector $m$ and diagonal scale $\diag[V]$ from Procedure \ref{alg:CI}; choice of uppermost tree level $\mathcal{T}_t$ for elicitation with $t \in \left\{0, \ldots, n-1\right\}$.}
	\KwResult{$n\times n$ positive definite correlation matrix $R$}
	Initialise $\Rho \leftarrow I$ with $I$ the $n\times n$ identity matrix \;
	\If{$t > 0$}{
		Select conditioning value $\hat{\mu}_1 \in \left\{a_1, b_1\right\}$ (Section \ref{sec:conditioning}) \;
		$\hat{\eta}_1 \leftarrow g(\hat{\mu}_1)$\;
		\For{$k = 2:n$}{
			\Repeat{expert accepts $f\left(\mu_k \mid m_{k \mid 1}, V_{k, k \mid 1}, r, s\right)$}{
				Elicit feasible $c_{k|1}$ for ${\mu}_{k} \mid u_k^\top, \hat{\mu}_{1}$ and set $m_{k|1} \leftarrow g(c_{k|1})$ (Section \ref{sec:estimation}) \;
				Parametrise $V_{k, 1}$ and $V_{k,k|1}$ from Eqs. \eqref{eq:Vkl} and \eqref{eq:Vkkl} \;
				Provide graphical and numerical feedback via Eqs. \eqref{eq:mucond_density}, \eqref{eq:mucond_dist} and \eqref{eq:mucond_q} \;
			}
			$\Rho_{1k} = \rho_{1, k} \leftarrow V_{k,1}/\sqrt{V_{1,1}V_{k,k}}$\;
		}
		\If{$t>1$}{
			\For{$l=2:t$}{
				evaluate $(a_{l|1:(l-1)}, b_{l|1:(l-1)})$ from $m_{l|1:(l-1)}$, $V_{l, l|1:(l-1)}$ (Section \ref{sec:conditioning}) \;
				select conditioning value $\hat{\mu}_{l} \in \left\{a_{l|1:(l-1)}, b_{l|1:(l-1)}\right\}$ \;
				$\hat{\eta}_l \leftarrow g\left(\hat{\mu_{l}}\right)$\;
				\For{$k=(l+1):n$}{
					\Repeat{expert accepts $f\left(\mu_k \mid m_{k \mid 1:l}, V_{k, k \mid 1:l}, r, s\right)$}{
						Elicit feasible $c_{k|1:l}$ for ${\mu}_{k} \mid u_k^\top, \hat{\mu}_{1:l}$ and set $m_{k|1:l} \leftarrow g(c_{k|1:l})$ (Section \ref{sec:estimation}) \;  
						Parametrise $V_{k, l}$ and $V_{k,k|1:l}$ from Eqs. \eqref{eq:Vkl} and \eqref{eq:Vkkl} \;
						Provide graphical and numerical feedback via Eqs. \eqref{eq:mucond_density}, \eqref{eq:mucond_dist} and \eqref{eq:mucond_q} \;
					}
					Evaluate $\Rho_{lk} = \rho_{l, k|1:(l - 1)}$ by Eq. \eqref{eq:cond_corr} with substitution $\gamma = l$ \;
				}
			}
		}
	}
	$R \leftarrow P$ by Algorithm \ref{alg:PR} \;
	\Return $R$.
	\caption{Elicitation of positive definite correlation matrix $R$.}\label{alg:cond}
\end{algorithm}

\begin{algorithm}[ht]
	\SetAlgoLined
	\KwData{Set of compound Poisson distributions $\mathcal{F}$ with $p \in (1, 2)$ and unknown dispersion; $s$ and $v_{\phi}(\hat{\mu}_0, w)$ from Procedure \ref{alg:dispersion};}
	\KwResult{ $r$ in Eq. \eqref{eq:model_lambda},  $p \in (1, 2)$;}
	Elicit $s$ and $v_{\phi}(\hat{\mu}_0, w)$ by Section \ref{sec:dispersion_EDM} and Procedure \ref{alg:dispersion} for sample mean $\bar{\psi} \mid \hat{\mu}_0, w$ \;
	\uIf{unknown $p \in (1, 2)$}{
		\Repeat{expert accepts $q_0 \sim UG(s, r_p)$, see Eq. \eqref{eq:UG}} {
			Elicit median for $q_0  = P(\psi = 0 \mid \hat{\mu}_0)$  to parametrise $r_p$ (Section \ref{sec:CP}) \;
			Parametrise $r$ and $p$ by Eq. \eqref{eq:UGparams} \;
			Provide graphical and numerical feedback \;
		}
	} 
	\Else{
		Given known $p \in (1, 2)$, set $r \leftarrow v_\phi(\hat{\mu}_0, w)ws/\hat{\mu}_0^p$ \;
	}
	\Return $s$, $r$, $p$.
	\caption{Elicitation of power parameter $p$ for compound Poisson model.}
	\label{alg:CP}
\end{algorithm}

\clearpage

\phantomsection

\addcontentsline{toc}{section}{References}

\bibliographystyle{chicago}
\bibliography{main}		

\end{document}